\documentclass[]{interact}

\usepackage{epstopdf}
\usepackage[caption=false]{subfig}

\usepackage[numbers,sort&compress]{natbib}
\bibpunct[, ]{[}{]}{,}{n}{,}{,}
\renewcommand\bibfont{\fontsize{10}{12}\selectfont}
\makeatletter
\def\NAT@def@citea{\def\@citea{\NAT@separator}}
\makeatother

\theoremstyle{plain}
\newtheorem{theorem}{Theorem}[section]

\theoremstyle{definition}
\newtheorem{definition}[theorem]{Definition}

\theoremstyle{remark}
\newtheorem{remark}{Remark}

\usepackage{latexsym,amsmath,amssymb,xspace,comment,graphicx, url, color}
\usepackage{comment}
\usepackage{algorithm}
\usepackage{algpseudocode}
\usepackage{here}
\usepackage{ulem}
\begin{document}


\title{\textcolor{red}{A search for short-period Tausworthe generators over $\mathbb{F}_b$ 
with application to Markov chain quasi-Monte Carlo}}

\author{
\name{Shin Harase\textsuperscript{a}\thanks{CONTACT Shin Harase. Email: harase@fc.ritsumei.ac.jp}}
\affil{\textsuperscript{a}
College of Science and Engineering, Ritsumeikan University, Shiga, Japan}
}

\maketitle

\begin{abstract}
A one-dimensional sequence $u_0, u_1, u_2, \ldots \in [0, 1)$ is said to be completely uniformly distributed (CUD) if overlapping $s$-blocks $(u_i, u_{i+1}, \ldots , u_{i+s-1})$, $i = 0, 1, 2, \ldots$, are uniformly distributed for every dimension $s \geq 1$. 
This concept naturally arises in Markov chain quasi-Monte Carlo (QMC). 
However, the definition of CUD sequences is not constructive, and 
thus there remains the problem of how to implement the Markov chain QMC algorithm in practice. 
Harase (2021) focused on the $t$-value, which is a measure of 
uniformity widely used in the study of QMC, and implemented 
short-period Tausworthe generators (i.e., linear feedback shift register generators) 
over the two-element field $\mathbb{F}_2$ 
that approximate CUD sequences by running for the entire period. 
In this paper, we generalize a search algorithm over 
$\mathbb{F}_2$ to that over arbitrary finite fields $\mathbb{F}_b$ with $b$ elements 
and conduct a search for Tausworthe generators over $\mathbb{F}_b$ 
with $t$-values zero (i.e., optimal) for dimension $s = 3$ and small for $s \geq 4$, 
especially in the case where $b = 3, 4$, and $5$. 
We provide a parameter table of Tausworthe generators over $\mathbb{F}_4$, 
and report a comparison between our new generators over $\mathbb{F}_4$ 
and existing generators over $\mathbb{F}_2$ 
in numerical examples using Markov chain QMC.
\end{abstract}

\begin{keywords}
Pseudorandom number generation; Quasi-Monte Carlo; Markov chain Monte Carlo; 
Bayesian inference; Linear regression
\end{keywords}

\section{Introduction}

We study the problem of calculating the expectation $E_{\pi}[f(\textcolor{red}{\mathbf{X}})]$ using Markov chain Monte Carlo (MCMC) methods for a target distribution $\pi$ \textcolor{red}{on a state space $\mathcal{X}$} 
and some function $f\textcolor{red}{: \mathcal{X} \to \mathbb{R}}$\textcolor{red}{, where $\mathbf{X}$ is a $\pi$-distributed random variable on $\mathcal{X}$}. 
We are interested in improving the accuracy by replacing 
IID uniform random points with quasi-Monte Carlo (QMC) points. 
However, traditional QMC points (e.g., Sobol', Niederreiter--Xing, 
Faure, and Halton) are not \textcolor{red}{straightforwardly} applicable. 
Motivated by a simulation study conducted by Liao \cite{doi:10.1080/10618600.1998.10474775}, 
Owen and Tribble~\cite{MR2168266} and Chen et al.~\cite{MR2816335} 
\textcolor{red}{theoretically} showed that 
Markov chain QMC remains consistent if the driving sequences are completely uniformly distributed (CUD). A one-dimensional sequence $u_0, u_1, u_2, \ldots$ $\in [0, 1)$ 
is said to be CUD if overlapping $s$-blocks $(u_i, u_{i+1}, \ldots , u_{i+s-1})$, $i = 0, 1, 2, \ldots$, are uniformly distributed for every dimension $s \geq 1$. 
Levin \cite{MR1731474} proposed some constructions of CUD sequences, but they are not suitable to implement. 
Thus, there remains the problem of how we implement the Markov chain QMC algorithm, 
in particular, how we construct suitable driving sequences in practice. 

Tribble and Owen \cite{MR2426105} and Tribble \cite{MR2710331}
proposed an implementation method to obtain point sets that approximate CUD sequences
by using short-period linear congruential and Tausworthe generators (i.e., linear feedback shift register generators 
over the two-element field $\mathbb{F}_2 :=  \{ 0, 1\}$) that run for the entire period. 
Moreover, Chen et al.~\cite{MR3173841} implemented short-period Tausworthe generators in terms of the equidistribution property, which is a coarse measure of uniformity 
in the area of pseudorandom number generation \cite{LP2009}. 

In \textcolor{red}{a} previous study, 
Harase~\cite{MR4143523} implemented short-period Tausworthe generators 
over $\mathbb{F}_2$ that approximate CUD sequences in terms of the $t$-value, 
which is a central measure in the theory of $(t, m, s)$-nets and $(t, s)$-sequences. 
The key technique was to use a polynomial analogue of Fibonacci numbers 
and their continued fraction expansion, which was originally proposed by Tezuka and Fushimi~\cite{MR1160278}. 
More precisely, we can view Tausworthe generators 
as a polynomial analogue of Korobov lattice rules 
with a denominator polynomial $p(x) \in \mathbb{F}_2[x]$ and a numerator polynomial $q(x) \in \mathbb{F}_2[x]$ (cf. \cite{MR2723077,MR1978160}), and 
hence, the $t$-value is zero (i.e., optimal) for dimension $s = 2$ 
if and only if the partial quotients in the continued fraction of $q(x)/p(x)$ are all of degree one \cite{MR1172997,MR1160278}. 
By enumerating such pairs of polynomials $(p(x), q(x))$ efficiently, 
Harase~\cite{MR4143523} conducted an exhaustive search of 
Tausworthe generators over $\mathbb{F}_2$ 
with $t$-values zero for $s = 2$ and small (but not zero) for $s \geq 3$, 
and demonstrated the effectiveness in numerical examples using Gibbs sampling.

From the theoretical and practical perspective, 
the most interesting case is the $t$-value zero. 
However, Kajiura et al.~\cite{MR3807854} proved that 
there exists no maximal-period Tausworthe generator over $\mathbb{F}_2$ 
with $t$-value zero for $s = 3$. 
In fact, in finite fields $\mathbb{F}_b$ of prime power order $b \geq 3$, 
we can find maximal-period Tausworthe generators with $t$-value zero 
for $s = 3$, for some combinations of $b$ and $m$. 

In this paper, our aim is to conduct an exhaustive search of 
maximal-period Tausworthe generators over $\mathbb{F}_b$ with $t$-values zero for  dimension $s = 3$, in addition to $s = 2$, especially in the case where $b = 3, 4$, and $5$. 
For this purpose, we generalize the search algorithms of Tezuka and Fushimi~\cite{MR1160278} and Harase~\cite{MR4143523} over $\mathbb{F}_2$ to those over $\mathbb{F}_b$. 
We provide a parameter table of Tausworthe generators over $\mathbb{F}_4$ with $t$-values zero for $s = 3$ and small for $s \geq 4$ to implement the Markov chain QMC algorithm. 
Accordingly, we report a comparison between our new Tausworthe generators over $\mathbb{F}_4$ 
and existing generators \cite{MR3173841,MR4143523} over $\mathbb{F}_2$ 
in numerical examples using Markov chain QMC. 

The rest of this paper is organized as follows: 
In Section~\ref{sec:preliminaries}, we recall the definitions of CUD sequences, Tausworthe generators, and the $t$-value, and recall a connection between the $t$-value and continued fraction expansion. 
In Section~\ref{sec:main}, we discuss our main results: 
In Section~\ref{subsec:multiplicity}, we investigate the number of polynomials $q(x)$ 
for which the partial quotients of the continued fraction expansion of $q(x)/p(x)$ all have degree one for a given irreducible polynomial $p(x)$ over $\mathbb{F}_b$. 
In Section~\ref{subsec:algorithm}, we describe a search algorithm 
of Tausworthe generators over $\mathbb{F}_b$. 
In Section~\ref{subsec:table}, we conduct an exhaustive search in the case where $b = 3, 4$, and $5$, and provide tables. 
In Section~\ref{sec:examples}, we present numerical examples, 
such as Gibbs sampling and a simulation of a queuing system, 
in which both Tausworthe generators over $\mathbb{F}_2$ and $\mathbb{F}_4$ 
optimized in terms of the $t$-value perform 
\textcolor{red}{comparable to or} better than 
Tausworthe generators \cite{MR3173841} optimized in terms of the equidistribution property. In Section~\ref{sec:conclusion}, we conclude this paper.

\section{Preliminaries}\label{sec:preliminaries}
We refer the reader to \cite{MR2683394,MR4143523,LP2009,MR2723077,MR1172997} for general information. 

\subsection{Discrepancy and completely uniformly distributed sequences} \label{subsec:cud}
 
Let $P_s = \{ \mathbf{u}_0,  \mathbf{u}_1, \ldots, \mathbf{u}_{N-1} \} \subset [0, 1)^s$ 
be an $s$-dimensional point set of $N$ elements in the sense of a \textit{multiset}. 
Let us recall the definition of discrepancy $D_N^{*s}(P_s)$ as a measure of uniformity.
\begin{definition}[Discrepancy] \label{def:discrepancy}
For a point set $P_s = \{ \mathbf{u}_0,  \mathbf{u}_1, \ldots, \mathbf{u}_{N-1} \} \subset [0,1)^s$, the (\textit{star}) \textit{discrepancy} is defined as 
\begin{eqnarray*}
 D_N^{*s} (P_s)  := \sup_{J} \left\lvert \frac{\nu (J; P_s)}{N} - {\rm vol}(J) \right\rvert,
 \end{eqnarray*}
where the supremum is taken over all intervals $J$ of the form $\prod_{j = 1}^s [0, t_j)$ 
for $0 < t_j \leq 1$, 
$\nu (J; P_s)$ denotes the number of $i$ with $0 \leq i \leq N-1$ 
for which $\mathbf{u}_i \in J$, 
and ${\rm vol}(J) := \prod_{j = 1}^{s} t_j$ denotes the volume of $J$. 
\end{definition}                                                                                                                                                                                                                                                                                                                                                                                                                                                                                                                                                                                                                                                                                                                                                                                                                                                                                                            
We define the CUD property 
for a one-dimensional sequence $\{ u_i \}_{i = 0}^{\infty}$ in $[0,1)$\textcolor{red}{, 
which is known as one of the definitions of random number sequences in \cite[Chapter~3.5]{MR3077153}}. 
\begin{definition}[CUD sequences] \label{def:cud} 
A one-dimensional infinite sequence $u_0, u_1,$ $u_2, \ldots$ $\in [0, 1)$ 
is said to be \textit{completely uniformly distributed} (\textit{CUD}) 
if overlapping $s$-blocks satisfy  
\begin{eqnarray*} \label{eqn:cud}
 \lim_{N \to \infty} D_N^{*s} \left( (u_0, \ldots, u_{s-1}), (u_1, \ldots, u_{s}), \ldots, (u_{N-1}, \ldots, u_{N+s-2}) \right) = 0
 \end{eqnarray*}
for every dimension $s \geq 1$; in short, 
the sequence of $s$-blocks $(u_i, \ldots, u_{i+s-1}),$ $i=0,1, \ldots$, is 
uniformly distributed in $[0,1)^s$ for every dimension $s \geq 1$. 
\end{definition}
In the study of Markov chain QMC, it is desirable that $D_N^{*s}$ 
converges to zero as fast as possible if $N \to \infty$ (cf. \cite{MR3275857,MR3563205}). 
As a necessary and sufficient condition of Definition~\ref{def:cud}, 
Chentsov \cite{CHENTSOV1967218} proved the following theorem:
\begin{theorem}[\cite{CHENTSOV1967218}] \label{thm:Chentsov}
A one-dimensional infinite sequence $u_0, u_1,$ $u_2, \ldots$ $\in [0, 1)$ is CUD if and only if 
non-overlapping $s$-blocks satisfy 
\begin{equation} \label{eqn:Chentsov}
 \lim_{N \to \infty} D_N^{*s} \left((u_0, \ldots, u_{s-1}), (u_{s}, \ldots, u_{2s-1}), \ldots, (u_{(N-1)s}, \ldots, u_{Ns-1}) \right) = 0
 \end{equation}
 for every dimension $s \geq 1$. 
 \end{theorem}
We thus use a sequence $\{ u_i \}_{i = 0}^{\infty}$ in $[0,1)$ for Markov chain QMC in this order. 

\subsection{Tausworthe generators over $\mathbb{F}_b$} \label{subsec:lcg}

Let $\mathbb{F}_b$ be a finite field with $b$ elements, where $b$ is a prime power,
and perform addition and multiplication over $\mathbb{F}_b$. 
\textcolor{red}{We define Tausworhe generators over $\mathbb{F}_b$, 
which are usually defined over $\mathbb{F}_2$ \cite{MR1325871,MR1620231,LP2009,MR0184406}. }

\begin{definition}[Tausworthe generators over $\mathbb{F}_b$] \label{def:Tausworthe}
Let $p(x) := x^m - c_1 x^{m-1} - \cdots - c_{m-1}x - c_m \in \mathbb{F}_b[x]$\textcolor{red}{, where $c_m \neq 0$}. 
We consider the linear recurrence \textcolor{red}{over $\mathbb{F}_b$ given by }
\begin{eqnarray} \label{eqn:m-sequence}
a_i  := c_1 a_{i-1} + \cdots + c_m a_{i-m} \in \mathbb{F}_b, \qquad \textcolor{red}{i = 0, 1, 2, \ldots, }
\end{eqnarray}
whose \textit{characteristic polynomial} is $p(x)$. 
Let $\sigma$ be a \textit{step size} and $w$ a \textit{digit number}. 
We define the output $u_i \in [0, 1)$ at step $i$ as 
\begin{eqnarray} \label{eqn:Tausworthe}
 u_i := \sum_{j = 0}^{w-1} \eta(a_{i \sigma +j}) b^{-j-1} \in [0,1),  \qquad i = 0, 1, 2, \ldots, 
 \end{eqnarray}
where $\eta: \mathbb{F}_b \to \mathbb{Z}_b := \{ 0, 1, \ldots, b-1\}$ is a bijection with $\eta (0) = 0$.
\textcolor{red}{If $p(x)$ is primitive, $(a_0, \ldots, a_{m-1}) \neq (0, \ldots, 0)$, $0 < \sigma < b^m-1$, 
and $\gcd (\sigma, b^m-1) = 1$, 
then the sequences (\ref{eqn:m-sequence}) and (\ref{eqn:Tausworthe}) are both purely periodic with maximal period $b^m-1$. 
Throughout this paper, we assume these maximal-period conditions.} 
We call a generator in such a class a \textit{Tausworthe generator over $\mathbb{F}_b$} 
(or a \textit{linear feedback shift register generator over $\mathbb{F}_b$}). 
\end{definition}

Similar to the case of $\mathbb{F}_2$, Tausworthe generators over $\mathbb{F}_b$ 
can be viewed as a polynomial analogue of \textit{linear congruential generators} (LCGs):
\begin{eqnarray*}
X_i(x) & := & q(x) X_{i-1}(x) \mod{p(x)} \label{eqn:LCG1},\\
X_i(x)/p(x) & = & a_{i \sigma} x^{-1} + a_{i \sigma+1}x^{-2} + a_{i \sigma+2}x^{-3} + \cdots \in \mathbb{F}_b((x^{-1})) 
\label{eqn:LCG2},
\end{eqnarray*}
where $X_i(x) \in \mathbb{F}_b[x], i = 0, 1, 2, \ldots$, is a sequence of polynomials, 
$p(x), q(x) \in \mathbb{F}_b[x]$ represent a \textit{modulus} and \textit{multiplier}, respectively, and 
the step size $\sigma$ satisfies $q(x) = x^{\sigma} \mod{p(x)}$ and $0 < \sigma < b^m-1$.
Then, the output $u_i$ in (\ref{eqn:Tausworthe}) is expressed as 
$u_i= \nu_w(X_i(x)/p(x))$, where a map $\nu_w: \mathbb{F}_b((x^{-1})) \to [0,1)$
is given by $\sum_{j = j_0}^{\infty} k_j x^{-j-1} \mapsto \sum_{j = \max{\{0,  j_0 \}}}^{w-1} \eta(k_j)b^{-j-1}$, 
which transforms a formal power series in $\mathbb{F}_b((x^{-1}))$ 
into a $b$-adic expansion with $w$ digits in $[0, 1)$. 

Moreover, similar to LCGs, Tausworthe generators have a lattice structure. 
Let $N := b^m$. We consider a sequence
\begin{eqnarray} \label{eqn:Tausworthe sequence}
u_0, u_1, \ldots, u_{N-2}, u_{N-1}= u_0, u_1, \ldots \in [0,1)
\end{eqnarray}
generated by a Tausworthe generator (\ref{eqn:Tausworthe})
with period length $N-1$. 
We set $s$-dimensional overlapping points $\mathbf{u}_i = (u_i, \ldots, u_{i+s-1})$ for $i=0,1, \ldots, N-2$, 
that is, $\mathbf{u}_0 = (u_0, \ldots, u_{s-1}), \mathbf{u}_1 = (u_1, \ldots,  u_s), \ldots, \mathbf{u}_{N-2} = (u_{N-2}, u_0, \ldots, u_{s-2})$.  
We construct a QMC point set
\begin{eqnarray} \label{eqn:point set}
P_s = \{ \mathbf{0} \} \cup \{ \mathbf{u}_i \}_{i = 0}^{N-2} \subset [0,1)^s,
\end{eqnarray}
adding the origin $\{ \mathbf{0} \}$. Note that the cardinality is $\# P_s = b^m$. 
Then, a point set $P_s$ in (\ref{eqn:point set}) can be viewed as a polynomial analogue of \textit{Korobov lattice rules}: 
\begin{eqnarray} \label{eqn:Korobov}
 P_s = \biggl\{ \nu_w \left(\frac{h(x)}{p(x)} (1, q(x), q(x)^2, \ldots, q(x)^{s-1}) \right) \Big\rvert
  \ \deg (h(x)) < m \biggl\},
\end{eqnarray}
where the map $\nu_w$ is applied component-wise and $m = \deg (p(x))$; see \cite{MR2723077,MR1978160} for details. 

A pair of polynomials $(p(x), q(x))$ is a parameter set of Tausworthe generators. 
Thus, in accordance with Definition~\ref{def:cud}, 
we would like to find a pair of polynomials $(p(x), q(x))$ 
with small discrepancy $D_N^{*s}(P_s)$ for each $s \geq 1$. 

\subsection{$t$-value and continued fraction expansion} \label{subsec:t-value}

A point set $P_s$ in (\ref{eqn:point set}) and (\ref{eqn:Korobov}) generated by  
a Tausworthe generator (\ref{eqn:Tausworthe}) is a \textit{digital net}. 
Hence, we can compute the $t$-value, which is closely related to $D_N^{*s}(P_s)$ for $N = b^m$.
\begin{definition}[$(t, m, s)$-nets] \label{def:(t, m, s)-net}
Let $s \geq 1$ and $0 \leq t \leq m$ denote integers. 
A point set $P_s$ of $b^m$ points in $[0, 1)^s$ 
is said to be a \textit{$(t, m, s)$-net} in base $b$ 
if every interval of the form $E = \prod_{j = 1}^s [ {r_j}/b^{d_j}, {(r_j +1)}/b^{d_j} )$ in $[0, 1)^s$ 
with integers $d_j \geq 0$ and $0 \leq r_j < b^{d_j}$ 
and of volume $b^{t-m}$ contains exactly $b^t$ points from $P_s$. 
\end{definition}
For a given dimension $s$, 
the smallest value $t$ for which $P_s$ is a $(t, m, s)$-net 
is said to be the \textit{$t$-value}. 
For a $(t, m, s)$-net $P_s$ in base $b$, 
we have an upper bound $D_N^{*s}(P_s) \leq C_{b, s} b^t(\log N)^{s-1}/N$, 
where the constant $C_{b, s} > 0$ only depends on $s$ and $b$; 
hence a small $t$-value is desirable. 
Therefore, we adopt the $t$-value as a measure of uniformity instead of the direct calculation of $D_N^{*s}(P_s)$ 
to obtain \textit{low-discrepancy} point sets.

Furthermore, in the case $s = 2$, 
there is a connection between the $t$-value of a polynomial Korobov lattice rule  (\ref{eqn:Korobov}) 
and the continued fraction expansion of $q(x)/p(x)$.  
Let $q(x)/p(x)$ be a rational function over $\mathbb{F}_b$ with $\gcd(p(x), q(x)) = 1$ and $\deg(p(x)) \geq 1$. 
Then, $q(x)/p(x)$ has a unique regular continued fraction expansion
\begin{eqnarray*}
q(x)/p(x) & = & A_{v+1}\textcolor{red}{(x)} + 1/(A_{v}(x)+ 1/(A_{v-1}(x)+\cdots+1/A_1(x))) \\
& = & [A_{v+1}(x); A_{v}(x), A_{v-1}(x), \ldots , A_1(x)]
\end{eqnarray*}
with a polynomial part $A_{v+1}(x) \in \mathbb{F}_b[x]$ and 
partial quotients $A_k(x) \in \mathbb{F}_b[x]$ satisfying $\deg(A_k(x)) \geq 1$ for $1 \leq  k \leq v$. 
Under this condition, we put  
$K(q/p) := \max_{1 \leq k \leq v} \deg(A_k(x))$. We have the following theorem:
\begin{theorem}[\cite{MR1172997,MR1160278}] \label{thm:Tezuka--Fushimi}
Let $p(x) \in \mathbb{F}_b[x]$ with $m = \deg (p(x))$. 
Let $q(x) \in \mathbb{F}_b[x]$ with $\deg (q (x)) < m$. 
Suppose that $\gcd (p(x),q(x)) = 1$. 
Then, the two-dimensional point set
\begin{eqnarray} \label{eqn:two-dim-lattice}
P_2 = \biggl\{  \nu_w \left( \frac{h(x)}{p(x)} (1, q(x)) \right) \Big\rvert \ \deg (h(x)) < m \biggl\}
\end{eqnarray}
is a $(t, m, 2)$-net in base $b$ with $t =K(q/p)-1$, which is exactly the $t$-value. 
In particular, $P_2$ has the $t$-value zero if and only if $K(q/p)=1$, 
so $\deg(A_k \textcolor{red}{(x)}) = 1$ for all $1 \leq k \leq v$ and $v = m$. 
\end{theorem}

Using the continued fraction expansion based on the above theorem, 
Tezuka and Fushimi \cite{MR1160278} proposed an algorithm to 
search for Tausworthe generators over $\mathbb{F}_2$ having pairs of polynomials $(p(x), q(x))$ with $t$-value zero for $s = 2$ and small for $s \geq 3$. 
Harase \cite{MR4143523} recently indicated that their technique 
is applicable to QMC points that approximate CUD sequences 
and conducted an exhaustive search over $\mathbb{F}_2$ removing some conditions. 

\textcolor{red}{
\begin{remark} \label{remark:L'Ecuyer}
In previous studies, L'Ecuyer and Lemieux \cite{10.1145/324138.324448,MR1978160} 
constructed short-period Tausworthe generators 
for QMC numerical integration in general-purpose use. 
To assess the uniformity of QMC points, 
they took into account the quality of the projections and 
developed several figures of merit 
using the equidistribution property, 
which are often used for selecting pseudorandom number generators 
with very long period \cite{MR1325871,MR1620231,10.1007/978-3-642-56046-0_21}. 
These figures of merit are implemented in 
LatNet Builder \cite{10.1007/978-3-030-98319-2_3}, a software tool 
to find good parameters, and are probably useful in our study, 
but they are not so closely related to 
the discrepancy $D_N^{*s}$ as the $t$-value 
because the condition of the equidistribution property 
is sometimes weaker than that of the $t$-value. 
The CUD sequences are defined via the discrepancy $D_N^{*s}$ in Definition~\ref{def:discrepancy}, 
so we adopt the $t$-value as a primary criterion. 
We also note that our study is aimed at an application to Markov chain QMC, 
not usual pseudorandom number generation. 
\end{remark}}



\section{Main results}\label{sec:main}

In the theory of $(t, m, s)$-nets and $(t, s)$-sequences, 
the most interesting case is the $t$-value zero. 
Kajiura et al.~\cite{MR3807854} proved that 
there exists no maximal-period Tausworthe generator over $\mathbb{F}_2$
with $t$-value zero for dimension $s = 3$. 
Thus, we conduct a search of Tausworthe generators over $\mathbb{F}_b$
with $t$-value zero for $s = 3$, 
especially in the case where $b = 3, 4$, and $5$. 

\subsection{Orthogonal multiplicity} \label{subsec:multiplicity}

To obtain a pair of polynomials $(p(x), q(x))$ with $t$-value zero for $s = 3$, 
it is necessary to satisfy at least $K(q/p) = 1$ in Theorem~\ref{thm:Tezuka--Fushimi}. 
Thus, we first investigate how many polynomials $q(x)$ 
satisfying $K(q/p) = 1$ exist for each irreducible polynomial $p(x) \in \mathbb{F}_b[x]$. 
For a given irreducible $p(x)$, 
we define the number
\begin{eqnarray*}
M(p) := \# \left\{ q(x) \in \mathbb{F}_b[x] \ \vert \ \deg (q(x)) < \deg(p(x)) \mbox{ and } K(q/p) = 1 \right\}. 
\end{eqnarray*}
The number $M(p)$ is called the \textit{orthogonal multiplicity} of $p(x)$ in \cite{MR1492892}. 
Specializing the proof 
for the case $\mathbb{F}_2$, Mesirov and Sweet \cite{MR909833} proved that  
every irreducible polynomial $p(x) \in \mathbb{F}_2[x]$ has exactly 
$M(p) = 2$ for $\deg(p(x)) \geq 2$\textcolor{red}{, that is, 
there exist only two polynomials $q(x)$ for which the partial quotients 
of $q(x)/p(x)$ have all degree one. 
Moreover, such polynomials are $q(x)$ and its inverse element $q^{-1}(x) \mod p(x)$, 
and hence, they yield exactly the same lattice point set $P_s$.}
This result asserts the existence of $P_2$ with $t$-value zero for every irreducible polynomial $p(x) \in \mathbb{F}_2[x]$ in Theorem~\ref{thm:Tezuka--Fushimi} 
\textcolor{red}{but also asserts that there is no degree of freedom to select such $q(x)$ for each $p(x)$.}

In fact, in the case $\mathbb{F}_b$ for $b \geq 3$, 
Blackburn \cite{MR1492892} indicated that the situation is different far from the case $\mathbb{F}_2$. 
More precisely, the orthogonal multiplicities $M(p)$ are not always the same number 
but are often much greater than two. 
Figure~\ref{fig:multiplicity} shows some histograms of orthogonal multiplicities $M(p)$ 
for all monic irreducible polynomials $p(x) \in \mathbb{F}_b[x]$ 
with $\deg(p(x)) = m$. 
No clear regularity as in $\mathbb{F}_2$ has been observed. 
Additionally, for arbitrary $\mathbb{F}_b$ with $b \geq 3$, 
it is not even known whether there exist irreducible polynomials 
$p(x) \in \mathbb{F}_b[x]$ with $M(p) = 0$ in general (see Remark~\ref{remark:Friesen}).
Thus, using computer calculations, we checked the existence of $M(p) > 0$  as follows:
\begin{figure}
\centering
\subfloat{\includegraphics[width=0.48\textwidth]{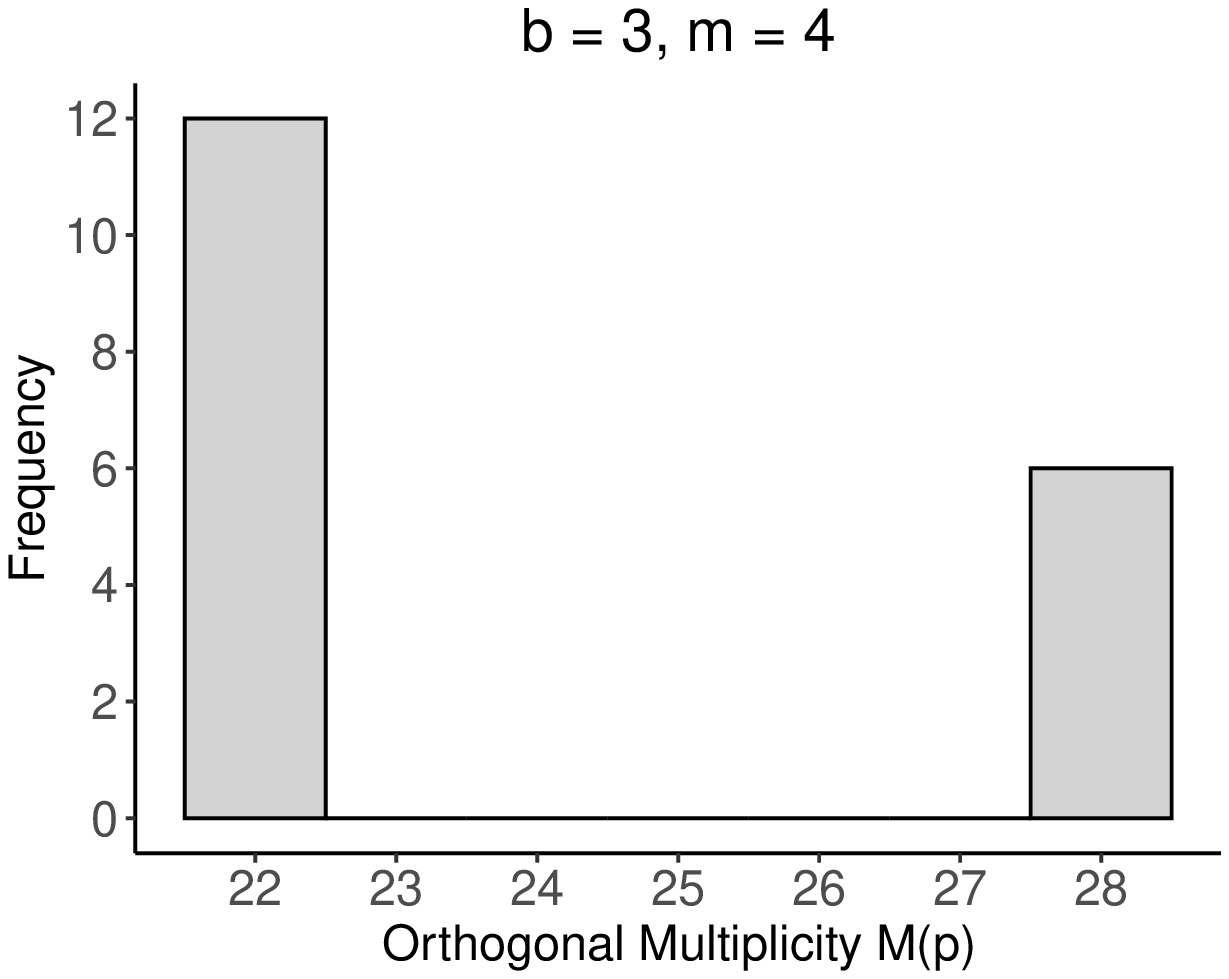}}
\subfloat{\includegraphics[width=0.48\textwidth]{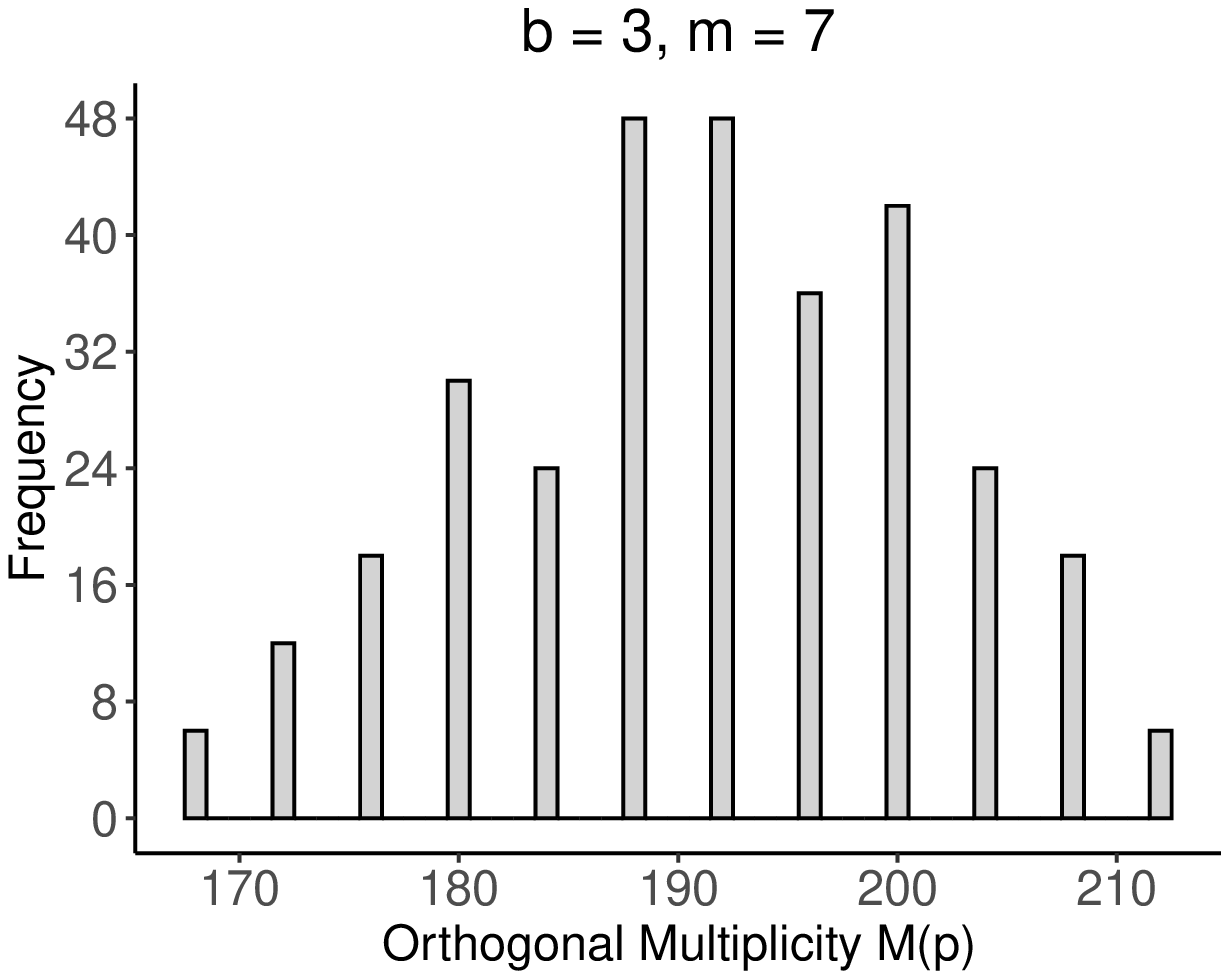}}\\
\subfloat{\includegraphics[width=0.48\textwidth]{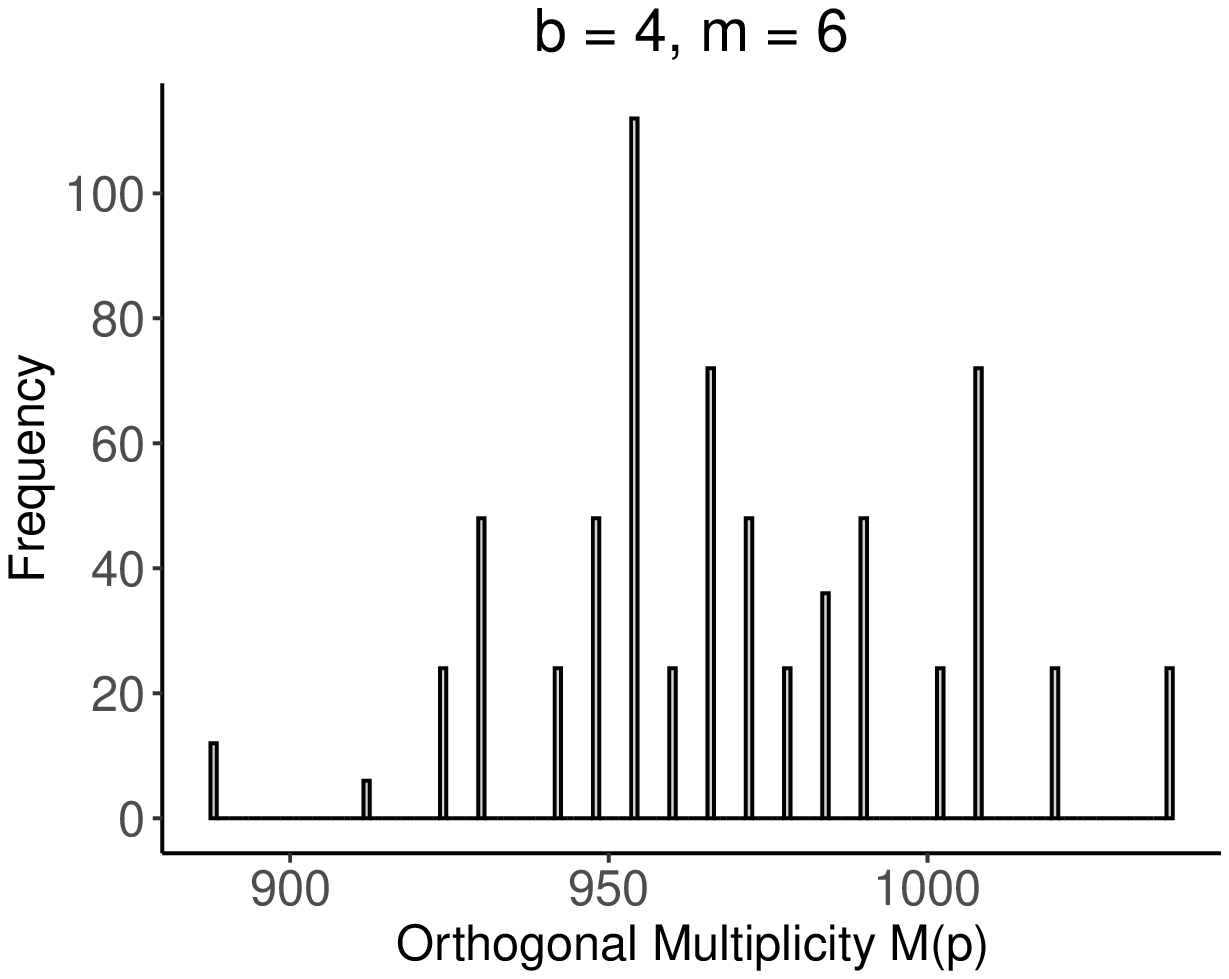}}
\subfloat{\includegraphics[width=0.48\textwidth]{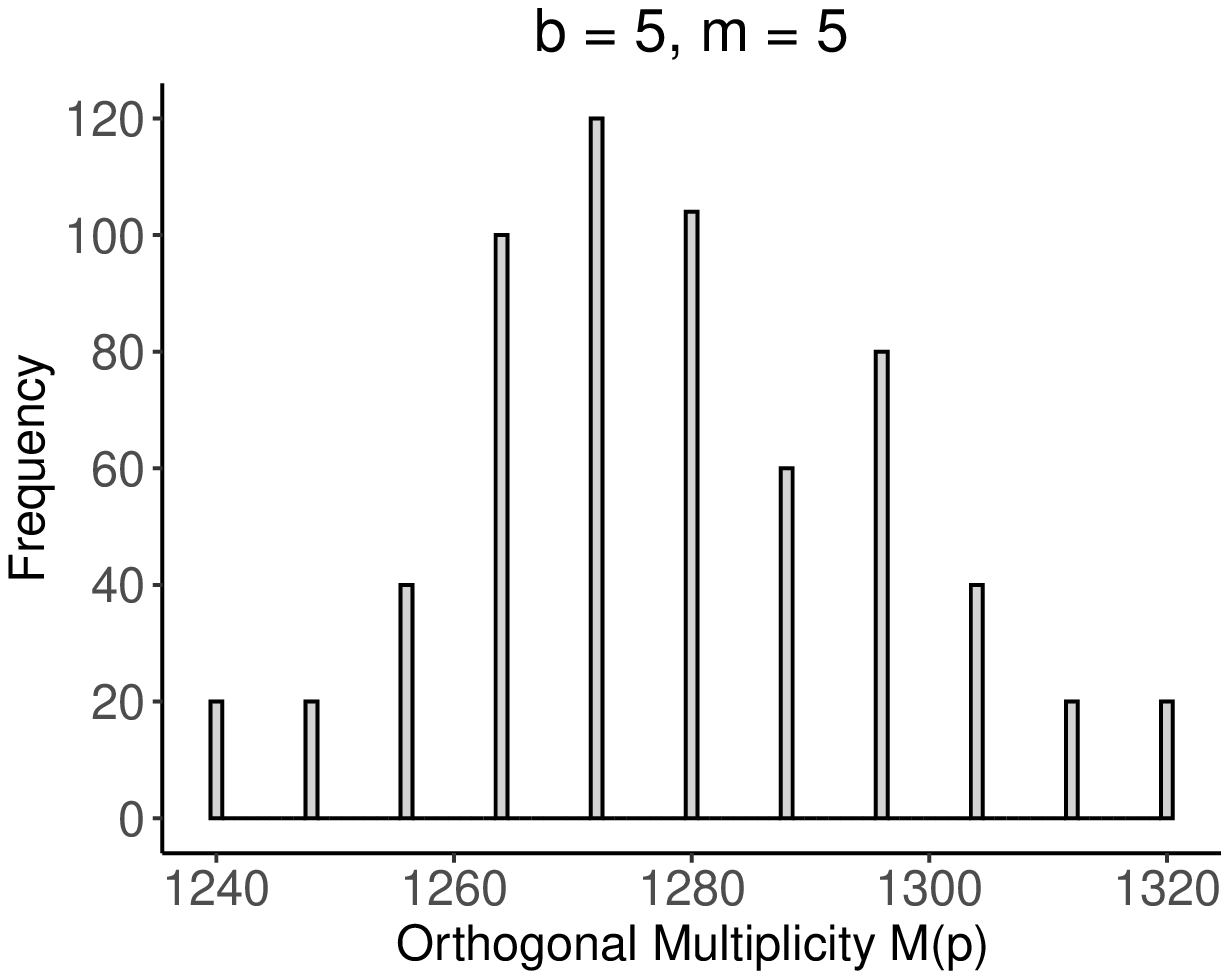}}
\caption{Distribution of orthogonal multiplicities $M(p)$ for all monic irreducible polynomials $p(x) \in \mathbb{F}_b[x]$ with $\deg(p\textcolor{red}{(x)})=m$.} \label{fig:multiplicity}
\end{figure}

\begin{theorem}
Let $\mathbb{F}_b$ be a finite field with $b$ elements. 
Every monic irreducible polynomial $p(x) \in \mathbb{F}_b$ with $\deg(p(x)) = m$ has $M(p) > 0$, at least under the following conditions: 
\begin{itemize} 
\item $1 \leq m \leq 15$ for $b = 3$;
\item $1 \leq m \leq 12$ for $b = 4$;
\item $1 \leq m \leq 10$ for $b = 5$.
\end{itemize}
\end{theorem}

\begin{remark} \label{remark:Friesen}
Let $p(x) \in \mathbb{F}_b[x]$ be an irreducible polynomial over $\mathbb{F}_b$. 
Assume that $0 < \deg (p(x)) < b$, that is, $\deg(p(x))$ is less than the order $b$ of $\mathbb{F}_b$. 
Under this condition, Friesen \cite[Theorem~2]{MR2354926} proved that 
every irreducible $p(x) \in \mathbb{F}_b[x]$ has $M(p) > 0$, 
that is, every irreducible $p(x)$ has $M(p)>0$  
provided the order $b$ of $\mathbb{F}_b$ is sufficiently large. 
This result is an improvement of that in the study by Blackburn \cite[Theorem~2]{MR1492892}. 
However, the assumption $\deg(p(x)) < b$ is significantly restrictive compared with the numerical results, 
and there has been no progress on the study of orthogonal multiplicities $M(p)$ since Friesen's study. 
Thus, we numerically checked the existence of $M(p) > 0$ only in the range required 
for our study.
\end{remark}

\subsection{A search algorithm using Fibonacci polynomials over $\mathbb{F}_b$} \label{subsec:algorithm}

Tausworthe generators associated with $(p(x), q(x))$ 
attain the $t$-value zero for $s = 3$ only if 
$K(q/p)=1$ in Theorem~\ref{thm:Tezuka--Fushimi}. 
Our strategy is to choose $(p(x), q(x))$ with $t$-value zero for $s = 3$ 
among pairs satisfying $K(q/p) = 1$. 
Thus, we generalize the search algorithms 
of Tezuka and Fushimi \cite{MR1160278} and Harase \cite{MR4143523} 
over $\mathbb{F}_2$ to those over arbitrary finite fields $\mathbb{F}_b$. 

\begin{figure}
\centering 
\includegraphics[width=0.9\textwidth]{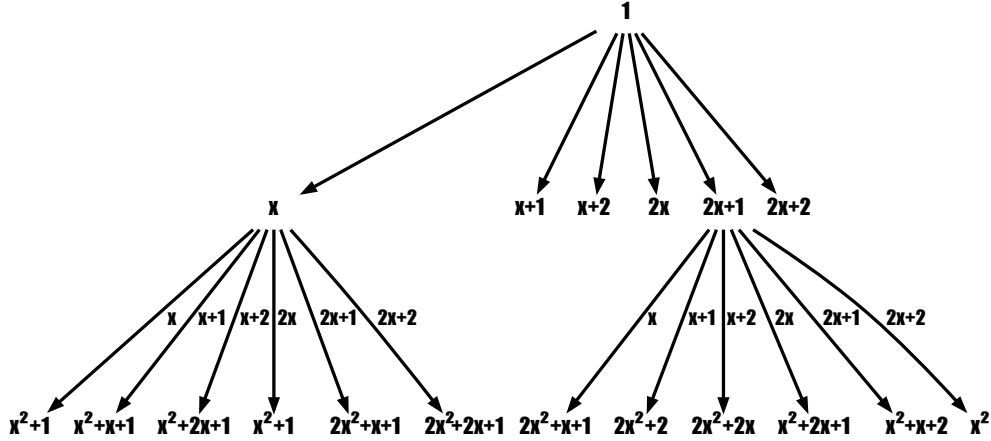}
\caption{Initial part of the tree of Fibonacci polynomials over $\mathbb{F}_3$.} \label{fig:tree}
\end{figure}

Recall that Fibonacci numbers $F_k, k = 1, 2 \ldots$, are defined by the recurrence $F_k = F_{k-1} + F_{k-2}$, 
where we choose the starting values $F_{-1} = 0, F_0 = 1$. 
Then, the continued fraction expansion of the ratio of two successive Fibonacci numbers $F_{k-1}/F_k$ is 
given by $F_{k-1}/F_{k} = [0; 1, 1, \ldots, 1]$ with partial quotients that are all one. 
As a polynomial analogue, we consider a sequence of polynomials $F_k(x), k = 1, 2, \ldots$, defined as
\begin{align}
& F_{k}(x)  =  A_{k}(x) F_{k-1}(x) + F_{k -2}(x) \label{eqn:Fibonacci1}\textcolor{red}{,} \\
& F_{-1}(x)  =  0, F_{0}(x)  =1 \label{eqn:Fibonacci2}\textcolor{red}{,} \\
& A_{k} (x)   =  \beta x + \gamma, \label{eqn:Fibonacci3} 
\end{align}
where $\beta \in \mathbb{F}_b^{*} \textcolor{red}{:= \mathbb{F}_b \backslash \{ 0 \}}$ and $\gamma \in \mathbb{F}_b$ so that $\deg (A_k\textcolor{red}{(x)}) = 1$. 
Similarly, we have the continued fraction expansion 
$F_{k-1}(x)/F_{k}(x) = [0; A_{k}(x), A_{k-1}(x), \ldots, A_{1}\textcolor{red}{(x)}]$, so $K(F_{k-1}/F_{k}) = 1$ holds.
The polynomials $F_k(x), k = 0, 1, 2, \ldots$, are called \textit{Fibonacci polynomials} over $\mathbb{F}_b$ (cf.~\cite{MR4231534}). 
Figure~\ref{fig:tree} shows an example of the initial part of a tree of Fibonacci polynomials 
over $\mathbb{F}_3$. Note that there exist $\{(b-1)b\}^m$ different pairs $(F_{m}(x), F_{m-1}(x))$ of Fibonacci polynomials over $\mathbb{F}_b$. 
Among pairs $(F_m(x), F_{m-1}(x))$,
we choose a suitable pair of $(p(x), q(x))$ 
with $t$-values zero for $s = 3$ 
and small for $s \geq 4$ satisfying Definition~\ref{def:Tausworthe}. 

We now generalize the algorithms \cite{MR4143523,MR1160278} 
over $\mathbb{F}_2$ to those over $\mathbb{F}_b$. 
Let $\textrm{lc}(F_m(x))$ denote the leading coefficient of $F_m(x)$ and 
$s_{\textrm{max}}$ denote a given maximum dimensionality. 
Our algorithm proceeds as follows: 
\begin{algorithm}[H]
\caption{Search algorithm}\label{algorithm:algo1}
\begin{algorithmic}[1]
\State Generate all the pairs ($(F_m(x), F_{m-1}(x)$) using the recurrence relation of 
Fibonacci polynomials (\ref{eqn:Fibonacci1})--(\ref{eqn:Fibonacci3}).
\State 
Set $F_{m}(x) \leftarrow F_{m}(x)/\textrm{lc}(F_m(x))$ 
and $F_{m-1}(x) \leftarrow F_{m-1}(x)/\textrm{lc}(F_m(x))$. 
\State Check the primitivity of $F_m(x)$.
\State Find $\sigma$ such that $x^{\sigma} = F_{m-1}(x) \mod F_{m}(x)$ and $0 < \sigma < b^m-1$. Check $\gcd (\sigma, b^m-1) = 1$. 
\State Choose pairs $(F_m(x),F_{m-1}(x))$ whose $t$-value is zero for $s = 3$. 
\State Compute the $t$-value $t^{(s)}$ for each dimension $s = 4, 5, \ldots, s_{\textrm{max}}$. 
For each $(F_m(x), F_{m-1}(x))$, 
construct a vector $(t^{(4)}, t^{(5)}, \ldots, t^{(s_{\textrm{max}})})$ of the $t$-values. 
\State Sort pairs $(F_m(x), F_{m-1}(x))$ in ascending order based on $(t^{(4)}, t^{(5)}, \ldots, t^{(s_{\textrm{max}})})$ starting from dimension $4$. 
\State Choose one of the best (or smallest) pairs $(F_m(x), F_{m-1}(x))$ in Step~7. 
\State Set $(p(x), q(x)) \gets (F_{m}(x), F_{m-1}(x))$. 
\end{algorithmic}
\end{algorithm}

Before we begin our algorithm, 
we create a lookup table of primitive polynomials over 
$\mathbb{F}_b$ in advance to avoid repeated computation in Step~3. 
In Step~2, $F_m(x)$ generated by (\ref{eqn:Fibonacci1}) is not always monic over arbitrary finite fields $\mathbb{F}_b$ except for $\mathbb{F}_2$, 
so it is necessary to divide $F_m(x)$ and $F_{m-1}(x)$ by the leading coefficient $\textrm{lc}(F_m(x))$. 
In Steps~5 and 6, we compute the $t$-values 
using Gaussian elimination \cite{Pirsic2001827}. 
For some combinations of $b$ and $m$, 
$(F_m(x), F_{m-1}(x))$ might not exist with $t$-value zero for $s = 3$ in Step~5. 
In this case, we skip Steps~6--9 and terminate the algorithm. 

\textcolor{red}{
\begin{remark} \label{remark:difference}
Tezuka and Fushimi \cite{MR1160278} and Harase \cite{MR4143523} dealt with 
the search algorithms that are similar to Algorithm~\ref{algorithm:algo1} 
but restricted to the special case $\mathbb{F}_2$. 
We now note that there are several differences between 
the cases $\mathbb{F}_2$ and $\mathbb{F}_b$ for $b \geq 3$. 
With regard to Equation~(\ref{eqn:Fibonacci3}), 
we have only two polynomials $A_k(x)$ with degree one in the case $\mathbb{F}_2$, 
that is, $A_k(x) = x$ or $x+1$; but we have many polynomials 
with degree one in general cases $\mathbb{F}_b$ (e.g., see Figure~\ref{fig:tree}). 
Thus, the patterns of continued fraction expansions 
$[0; A_{k}(x), A_{k-1}(x), \ldots, A_{1}(x)]$ drastically increase 
as opposite to the case $\mathbb{F}_2$. 
Moreover, every polynomial over $\mathbb{F}_2$ is always monic, 
and hence, Step~2 in Algorithm~\ref{algorithm:algo1} is not appeared in the existing algorithms. 
Once again, as mentioned in Section~3.1, there are only two 
polynomials $q(x)$ for which the partial quotients of $q(x)/p(x)$ have all degree one over $\mathbb{F}_2$, 
but many polynomials $q(x) $ with such property exist in the case $\mathbb{F}_b$. 
Therefore, our generalization would not be straightforward and simple 
when we search for parameters in practice.
\end{remark}}

\subsection{Specific parameters} \label{subsec:table}

We conduct an exhaustive search of short-period Tausworthe generators  
over $\mathbb{F}_3$, $\mathbb{F}_4$, and $\mathbb{F}_5$ using Algorithm~\ref{algorithm:algo1}. 
We set $s_{\textrm{max}} = 20$. 
If $b$ is a prime number (i.e, $b = 3$ or $5$), 
we identify $\mathbb{F}_b$ with $\mathbb{Z}_b$ and set a bijection $\eta: \mathbb{F}_b \to \mathbb{Z}_b$ as the identity map. 
If $b=4$, we set $\mathbb{F}_4 = \{ 0, 1,\alpha, \alpha^2 \}$ with $\alpha^2 = \alpha+1$ and $\alpha^3 = 1$ and set a bijection $\eta: \mathbb{F}_4 \to \mathbb{Z}_4 $ consisting of 
\[ 0 \mapsto 0, 1 \mapsto 1, \alpha \mapsto 2, \alpha^2=\alpha+1 \mapsto  3. \] 
Table~\ref{table:numbers} summarizes the number of maximal-period Tausworthe generators 
with $t$-value zero for dimension $s = 3$. 
We observe that a very few pairs of polynomials $(p(x), q(x))$ exist over $\mathbb{F}_3$; 
however, many pairs exist over $\mathbb{F}_4$ and $\mathbb{F}_5$, 
at least within the range described in the table. 
From the viewpoint of applications, 
we tabulate specific parameters of pairs of polynomials 
$(p(x), q(x))$ over $\mathbb{F}_4$ and step sizes $\sigma$ for $2 \leq m \leq 11$ in Table~\ref{table:parameters}.
In Table~\ref{table:parameters}, each first and second row shows the coefficients of $p(x)$ and $q(x)$ respectively; for example, $\alpha^2 \ 1 \ 1$ means $\alpha^2 + x + x^2$. 
Table~\ref{table:t-values} shows the $t$-values in the range of $1 \leq s \leq 20$. 
Throughout our search, we find several parameters 
with the same $t$-values, so we choose one from them. 

\begin{table}[H]
\caption{Number of pairs of polynomials $(p(x), q(x))$ that attain  
maximal-period Tausworthe generators with $t$-value zero for dimension $s = 3$.} 
\label{table:numbers}

Number of Tausworthe generators over $\mathbb{F}_3$ with $t$-value zero.\\
\begin{tabular}{c|rrrrrrrrrrrr} \hline
$m$ & 2 & 3 & 4 & 5 & 6 & 7 & 8 & 9 & 10 & 11 & 12 & 13\\ \hline
Num. & 8 & 6 & 0 & 0 & 8 & 6 & 0 & 0 & 0 & 0 & 0 & 0\\ \hline
\end{tabular} 
\\

Number of Tausworthe generators \textcolor{red}{over $\mathbb{F}_4$} 
with $t$-value zero.\\
\begin{tabular}{c|rrrrrrrrrr} \hline
$m$ & 2 & 3 & 4 & 5 & 6 & 7 & 8 & 9 & 10  & 11\\ \hline
Num. & 32 & 72 & 128 & 1296 & 2016 & 7648 & 4640 & 5328 & 4176 & 4560\\ \hline
\end{tabular} 
\\

Number of Tausworthe generators \textcolor{red}{over $\mathbb{F}_5$} with $t$-value zero.\\
\begin{tabular}{c|rrrrrrr} \hline
$m$ & 2 & 3 & 4 & 5 & 6 & 7 & 8 \\ \hline
Num. & 32 & 480 & 1056 & 16800 & 38720 & 514640 & 706496 \\ \hline
\end{tabular}
\end{table}

\begin{table}[H]
\caption{Specific parameters of pairs of polynomials $(p(x), q(x))$ over $\mathbb{F}_4$ 
and step sizes $\sigma$. } \label{table:parameters}
\begin{tabular}{l|l} \hline 

$m = 2$ & $\alpha^2 \ 1 \ 1$ \\ 
{} & $\alpha \ 1$ \quad ($\sigma = 8$) \\ \hline 

$m = 3$ & $\alpha^2 \ \alpha^2 \ \alpha^2 \ 1$ \\ 
{} & $1 \ \alpha \ \alpha^2$ \quad ($\sigma = 47$) \\ \hline 

$m = 4$ & $\alpha^2 \ \alpha^2 \ \alpha^2 \ 0 \ 1$ \\ 
{} & $\alpha^2 \ 1 \ 1 \ \alpha^2$ \quad ($\sigma = 131$) \\ \hline   

$m = 5$ & $\alpha^2 \ \alpha^2 \ \alpha \ 1 \ 0 \ 1$ \\ 
{} &  $\alpha \ \alpha^2 \ \alpha^2 \ \alpha^2 \ \alpha^2$ \quad ($\sigma = 724$)  \\ \hline  

$m = 6$ & $\alpha^2 \ 1 \ 0 \ 1 \ 1 \ 0 \ 1$ \\ 
{} & $1 \ 1 \ \alpha^2 \ \alpha^2 \ 1 \ \alpha$ \quad ($\sigma = 2267$) \\ \hline


$m = 7$ & $\alpha \ \alpha^2 \ 0 \ \alpha \ \alpha^2 \ \alpha \ \alpha \ 1$ \\ 
{} & $0 \ 0 \ \alpha^2 \ \alpha^2 \ \alpha \ \alpha^2 \ 1$ \quad ($\sigma = 1633$) \\ \hline  

$m = 8$ & $\alpha \ \alpha^2 \ 1 \ 1 \ 0 \ \alpha \ 0 \ 0 \ 1$ \\ 
{} & $1 \ 1 \ 1 \ 1 \ 0 \ 0 \ \alpha \ \alpha^2$ \quad ($\sigma = 16423$) \\ \hline  

$m = 9$ & $\alpha^2 \ \alpha^2 \ \alpha \ 0 \ 1 \ \alpha \ \alpha \ 1 \ 0 \ 1$ \\ 
{} & $\alpha \ 1 \ 1 \ \alpha^2 \ \alpha^2 \ \alpha^2 \ \alpha \ 0 \ 1$ \quad ($\sigma = 36887$) \\ \hline  

$m = 10$ & $\alpha \ \alpha^2 \ \alpha \ 0 \ 1 \ \alpha^2 \ 0 \ 0 \ \alpha^2 \ 0 \ 1$ \\
{} & $\alpha^2 \ 0 \ 0 \ \alpha \ 1 \ 0 \ 1 \ 1 \ 1 \ 1$ \quad ($\sigma = 1030108$) \\ \hline 
$m = 11$ & $\alpha^2$ \ $\alpha$ \ 1 \ $\alpha^2$ \ $\alpha$ \ $\alpha^2$ \ 1 \ $\alpha^2$ \ $\alpha^2$ \ 1 \ $\alpha$ \ 1 \\ 
{} & $\alpha^2$ \ $\alpha$ \ $\alpha^2$ \ $\alpha$ \ $\alpha$ \ $\alpha^2$ \ 1 \ $\alpha^2$ \ 1 \ 1 \ $\alpha$ \quad ($\sigma = 3144209$)\\ \hline
 \end{tabular}
\end{table}

\begin{table}[H]
\caption{The $t$-values for good Tausworthe generators over $\mathbb{F}_4$ .} \label{table:t-values}
{\footnotesize
\begin{tabular*}{15cm}{@{\extracolsep{\fill}}c|rrrrrrrrrrrrrrrrrrrrr} \hline
$m \backslash s$ & $1$ & 2 & 3 & 4 & 5 & 6 & 7 & 8 & 9 & 10 & 11 & 12 & 13 & 14 & 15 & 16 & 17 & 18 & 19 & 20 \\ \hline \hline
$ 2$ & 0 & 0 & 0 & 0 & 0 & 1 & 1 & 1 & 1 & 1 & 1 & 1 & 1 & 1 & 1 & 1 & 1 & 1 & 1 & 1 \\ \hline
$3$ & 0 & 0 & 0 & 1 & 1 & 1 & 1 & 1 & 1 & 1 & 1 & 1 & 1 & 1 & 1 & 1 & 1 & 1 & 1 & 1 \\ \hline
$4$ & 0 & 0 & 0 & 1 & 1 & 1 & 1 & 2 & 2 & 2 & 2 & 2 & 2 & 2 & 2 & 2 & 2 & 2 & 2 & 2 \\ \hline
$5$ & 0 & 0 & 0 & 1 & 1 & 2 & 2 & 2 & 2 & 2 & 2 & 2 & 2 & 2 & 2 & 2 & 2 & 2 & 2 & 2 \\ \hline
$6$ & 0 & 0 & 0 & 1 & 2 & 2 & 2 & 2 & 3 & 3 & 3 & 3 & 3 & 3 & 3 & 3 & 3 & 3 & 3 & 3 \\ \hline
$7$ & 0 & 0 & 0 & 1 & 2 & 2 & 2 & 3 & 3 & 3 & 3 & 3 & 3 & 4 & 4 & 4 & 4 & 4 & 4 & 4 \\ \hline
$8$ & 0 & 0 & 0 & 1 & 2 & 4 & 4 & 4 & 4 & 4 & 4 & 4 & 4 & 4 & 4 & 4 & 4 & 4 & 4 & 4 \\ \hline
$9$ & 0 & 0 & 0 & 1 & 3 & 3 & 3 & 3 & 3 & 4 & 4 & 4 & 4 & 4 & 4 & 5 & 5 & 5 & 5 & 5 \\ \hline
$10$ & 0 & 0 & 0 & 2 & 2 & 3 & 3 & 3 & 4 & 4 & 4 & 5 & 5 & 6 & 6 & 6 & 6 & 6 & 6 & 6 \\\hline
$11$ & 0 & 0 & 0 & 2 & 3 & 3 & 3 & 4 & 5 & 5 & 5 & 5 & 5 & 5 & 5 & 5 & 5 & 6 & 6 & 6 \\ \hline 
\end{tabular*}}
\end{table}

\textcolor{red}{For the implementation, w}e introduce a \textcolor{red}{reasonably fast algorithm} to generate the output values (\ref{eqn:Tausworthe}) 
from Tausworthe generators over $\mathbb{F}_4$. 
Assume that $m \leq w$.
Let $\tilde{\mathbf{x}}_i = (a_{i \sigma}, a_{i \sigma +1}, \ldots, a_{i \sigma +m-1}, a_{i \sigma + m}, \ldots, a_{i \sigma + w-1})^\top \in \mathbb{F}_b^w$ denote a state vector at step $i$ 
\textcolor{red}{(${}^\top$ means ``transposed'')}.
We can define a state-space representation 
$\tilde{\mathbf{x}}_{i + 1} = \tilde{\mathbf{B}} \tilde{\mathbf{x}}_{i}$, 
where 
\begin{eqnarray} \label{eqn:transition}
 \tilde{\mathbf{B}} = \begin{pmatrix}
 \tilde{\mathbf{b}}_0 & \tilde{\mathbf{b}}_1& \cdots & \tilde{\mathbf{b}}_{m-1} & \mathbf{0} & \cdots & \mathbf{0}
\end{pmatrix}
\end{eqnarray}
is a $w \times w$ state transition matrix in $\mathbb{F}_b$
that consists 
of $m$ column vectors $\tilde{\mathbf{b}}_0, \tilde{\mathbf{b}}_1, \ldots,$ $\tilde{\mathbf{b}}_{m-1} \in \mathbb{F}_b^w$ 
and $w-m$ zero column vectors $\mathbf{0} \in \mathbb{F}_b^w$.
We now set $b = 4$ and decompose $a_i \in \mathbb{F}_4$, $i = 0, 1, \ldots$, into $a_i = \overline{a}_i \alpha + \underline{a}_i$ 
with $\overline{a}_i, \underline{a}_i \in \mathbb{F}_2$ \textcolor{red}{since a set $\{1, \alpha \}$ is a basis of $\mathbb{F}_4$ over $\mathbb{F}_2$}. Then, we can write 
$\mathbf{x}_{i + 1} =  \overline{a}_{i \sigma} (\alpha \tilde{\mathbf{b}}_0) + \underline{a}_{i \sigma} \tilde{\mathbf{b}}_0 + 
\overline{a}_{i \sigma +1} (\alpha \tilde{\mathbf{b}}_1) + \underline{a}_{i \sigma +1} \tilde{\mathbf{b}}_1 
+ \cdots + \overline{a}_{i \sigma +m-1} (\alpha \tilde{\mathbf{b}}_{m-1})+ \underline{a}_{i \sigma +m-1} \tilde{\mathbf{b}}_{m-1}$, that is, a linear combination of $2m$ column vectors 
$\alpha \tilde{\mathbf{b}}_{j}, \tilde{\mathbf{b}}_{j} \in \mathbb{F}_4^{w}$, $j = 0, 1, \ldots, m-1$, 
with coefficients in $\mathbb{F}_2$. 
From this, we can calculate $\tilde{\mathbf{x}}_i$ by only adding vectors $\alpha \tilde{\mathbf{b}}_{j}$ if $\overline{a}_{i\sigma+j} = 1$ and $\tilde{\mathbf{b}}_{j}$ if $\underline{a}_{i \sigma + j} = 1$ for each $j$. 
Moreover, the elements $0, 1, \alpha, \alpha +1 \in \mathbb{F}_4$ can be 
represented as column vectors 
$(0 \ 0)^\top, (0 \ 1)^\top, (1 \ 0)^\top, (1 \ 1)^\top \in \mathbb{F}_2^2$, respectively, 
and hence, $\alpha \tilde{\mathbf{b}}_j, \tilde{\mathbf{b}}_j, \tilde{\mathbf{x}}_{i+1} \in \mathbb{F}_4^{w}$ can be viewed as column vectors in $\mathbb{F}_2^{2w}$. Using this property, we can generate $\{ u_i \}_{i = 0}^{\infty}$ in (\ref{eqn:Tausworthe}) with reasonable speed, as if we performed additions over $\mathbb{F}_2$. The sample code is available at \url{https://github.com/sharase/cud-f4}.

\begin{remark} \label{remark:f2-lin}
Kajiura et al.~\cite{MR3807854} proved that 
there exists no Tausworthe generator over $\mathbb{F}_2$ 
with both maximal periodicity and $t$-value zero for $s = 3$ if $m \geq 3$. 
More precisely, they proved that $\mathbb{F}_2$-linear generators, 
which are a general class of linear pseudorandom number generators over $\mathbb{F}_2$
including Tausworthe generators (cf. \cite{LP2009,MR2723077}), 
have the $t$-value zero for $s = 3$ 
only if the period length is exactly three. 
Their proof was specialized for the case $\mathbb{F}_2$; for example, 
they used the property $\mathcal{L}_m \cap \mathcal{U}_m = \{ \mathbf{I}_{\textcolor{red}{m}} \}$ in \cite[Proof of Theorem~1]{MR3807854}, where \textcolor{red}{$\mathbf{I}_m$ 
denotes the identity matrix of order $m$, and} 
$\mathcal{L}_m$ and $\mathcal{U}_m$ denote 
a set of non-singular $m \times m$ lower-triangular 
and upper-triangular matrices, respectively.
This is false in the fields $\mathbb{F}_b$ except for $\mathbb{F}_2$. 
Indeed, Harase \cite{MR4143523} obtained 
Tausworthe generators over $\mathbb{F}_2$ 
with $t$-value two or three for $s = 3$, 
but they were not optimal with respect to the $t$-value. 
Thus, we conducted a search over $\mathbb{F}_b$, 
whose restrictions are looser than those over $\mathbb{F}_2$. 
\end{remark}

\begin{remark} \label{remark:transition}
\textcolor{red}{
We consider a reason why there are a very few pairs of polynomials $(p(x), q(x))$ 
with the $t$-value zero for $s = 3$ over $\mathbb{F}_3$ in Table~\ref{table:numbers}. 
Assume that $m \leq w$.
Let $\mathbf{A}_0$ denote the $(m \times m)$-transpose companion matrix 
of $p(x)$ in $\mathbb{F}_b$ given by 
\[ \mathbf{A}_0 = 
\begin{pmatrix}
{} & 1 & {} & {}\\
{} & {} & \ddots & {} \\
{} & {} & {} & 1\\
c_m & c_{m-1} & \cdots & c_1
\end{pmatrix}, \]
where blank entries in this matrix mean zeros. 
We set an $(m \times m)$-matrix $\mathbf{A} = \mathbf{A}_0^\sigma$. 
According to \cite[\S~5.1]{LP2009}, 
we can obtain the state transition matrix $\tilde{\mathbf{B}}$ in (\ref{eqn:transition})
 by expanding $\mathbf{A}$, that is, 
 if $m = w$, then we put $\tilde{\mathbf{B}} = \mathbf{A}$, and if $m < w$, for $j = m+1, \ldots, w$, 
we attach $(c_1^{(j)}, \ldots, c_{m}^{(j)})$ as the $j$th row vector, 
where the coefficients $c_i^{(j)}$ are given by the relation $a_{i \sigma+j} = c_{1}^{(j)} a_{i-1} + \cdots +c_{m}^{(j)} a_{i-m}$, 
and add $w-m$ columns of the zero vector $\mathbf{0}$. 
Thus, the $t$-value for dimension $s$ is determined by the 
the maximum number of linear independence 
of leading row vectors of $s$ generating matrices 
($\mathbf{I}_m$, $\mathbf{A}, \ldots, \mathbf{A}^{s-1})$; see \cite[Theorem~4.28]{MR1172997} or \cite[Theorem~4.52]{MR2683394} for details. 
In our construction scheme, one can only change the parameter values $c_1, \ldots, c_m$ and $\sigma$, 
so that the search space is restricted.}

\textcolor{red}{As an alternative, we have conducted a numerical experiment for which 
we discard the structure of Tausworthe generators 
and take general $(m \times m)$-full rank matrices $\mathbf{A}$, not given by 
$\mathbf{A}_0^{\sigma}$, 
as described in \cite[Equ.~(4)]{MR3807854}. 
(Here, we may assume without loss of generality that the row vectors 
$(c_1^{(j)}, \ldots, c_{m}^{(j)})$ are arbitrary.) 
Our goal here is to find a full rank matrix $\mathbf{A}$ such that 
a digital net generated by $(\mathbf{I}, \mathbf{A}, \mathbf{A}^2)$ 
has the $t$-value zero for $s = 3$ and the multiplicative order of $\mathbf{A}$ is $b^m-1$. 
For this, we generate full rank matrices $\mathbf{A}$ at random and check the above conditions. 
In computer search, we have confirmed the existence of such $\mathbf{A}$ in $\mathbb{F}_3$ for $2 \leq m \leq 10$. 
It might be expected that the existence holds true for every $m$ in arbitrary $\mathbb{F}_b$ except for $\mathbb{F}_2$. 
However, this approach seems to be significantly inefficient and time-consuming 
because it is not so easy to find matrices $\mathbf{A}$ 
that generate the digital nets with the $t$-value zero even for $s = 2$ if $m$ is large for small $b$. 
Therefore, it would be desirable to design 
some mathematical structure of $\mathbf{A}$ in advance 
before we conduct a search. 
In contrast, our algorithm always ensures the $t$-value zero for $s=2$. 
In this paper, we conduct an exhaustive search, 
but our algorithm has the advantage that 
we can easily switch from an exhaustive search to 
a random search by generating $A_k(x)$ in (\ref{eqn:Fibonacci3}) randomly. }
\end{remark}

\section{Numerical examples}\label{sec:examples}

We provide numerical examples to confirm the performance of Markov chain QMC. 
In our examples, we estimate the expectation $E_{\pi} [f(\mathbf{X})]$ and  
compare the following driving sequences:
\begin{enumerate}
\item[(a)] New: our new Tausworthe generators over $\mathbb{F}_4$;
\item[(b)] Harase: Tausworthe generators over $\mathbb{F}_2$ developed by Harase \cite{MR4143523};
\item[(c)] Chen: Tausworthe generators over $\mathbb{F}_2$ developed by Chen et al.~\cite{MR3173841}; and
\item[(d)] IID: Mersenne Twister \cite{Matsumoto:1998:MTE:272991.272995}. 
\end{enumerate}

We briefly explain how to use Tausworthe generators over $\mathbb{F}_b$.  
Recall that $N = b^m$ and the period length is $N-1$. 
For the output values (\ref{eqn:Tausworthe}) generated by Tausworthe generators, 
if $\textrm{gcd}(s, N-1) = 1$, 
we simply define $s$-dimensional non-overlapping points starting from the origin:
\begin{eqnarray} \label{eqn:gibbs_input} 
(0,  \ldots, 0), (u_0, \ldots, u_{s-1}), (u_s, \ldots, u_{2s-1}), \ldots, (u_{(N-2)s}, \ldots, u_{(N-1)s-1}). 
\end{eqnarray}
If $\textrm{gcd}(s, N-1) = d > 1$, instead of (\ref{eqn:gibbs_input}), 
we generate $d$ distinct short loops of $s$-dimensional points, that is, 
\begin{equation} \label{eqn:blocks}
(u_{j}, \ldots, u_{j+s-1}), (u_{j+s}, \ldots, u_{j+2s-1}), \ldots, (u_{j+ (((N-1)/d)-1)s}, \ldots, u_{j + ((N-1)/d)s -1}), 
\end{equation} 
for $j = 0, \ldots, d-1$, and concatenate them starting from the origin $(0, \ldots, 0)$ in this order. 
For these points, we apply $b$-adic digital shifts, that is, we add $(z_1, \ldots, z_s)$ to each 
$s$-dimensional point using the digit-wise addition $\oplus_{b}$ (see Remark~\ref{remark:digital_shift}), 
where $z\label{remark:digital_shift}_1, \ldots, z_s$ are IID samples from \textcolor{red}{$\mathcal{U}(0,1)$, 
that is, the continuous uniform distribution over $(0, 1)$}. 
We use the resulting points as input for Markov chain QMC; 
see Remark~\ref{remark:skips} and \cite{ChenThesis,MR2168266,MR2710331,MR2426105} for more details. 
We set $w = 32$ over $\mathbb{F}_2$ and $w = 16$ over $\mathbb{F}_4$ as a digit number in Definition~\ref{def:Tausworthe}. 
\begin{remark} \label{remark:digital_shift}
We recall the definition of \textit{digital shifts}. 
For $x = \sum_{j = 0}^{\infty} \xi_j b^{-j-1} \in [0, 1)$ and 
$z = \sum_{j = 0}^{\infty} \zeta_j b^{-j-1} \in (0,1)$ 
with $\xi_j, \zeta_j \in \mathbb{Z}_b$, 
we define the $b$-adic \textit{digitally shifted point} $\tilde{x} \in (0,1)$ 
as $\tilde{x} = x \oplus_b z := \sum_{j = 0}^{\infty} \psi_j b^{-j-1}$, 
where $\psi_j := \eta(\eta^{-1}(\xi_j) + \eta^{-1}(\zeta_j))$ with 
$\psi_j \neq b-1$ 
for infinitely many $j$ and `$+$' represents the addition in $\mathbb{F}_b$. 
For higher dimensions $s > 1$, let $\mathbf{z} = (z_1, \ldots, z_s) \in (0,1)^s$. 
For $\mathbf{x} = (x_{1}, \ldots, x_{s}) $, 
we similarly define the $b$-adic \textit{digitally shifted point} $\tilde{\mathbf{x}} \in (0, 1)^s$ 
as $\tilde{\mathbf{x}} = \mathbf{x} \oplus_b \mathbf{z} := (x_1 \oplus_b z_1, \ldots, x_s \oplus_b z_s)$.
\end{remark}

\subsection{Gaussian Gibbs sampling}

Our first example is a systematic Gibbs sampling scheme to generate 
the $s$-dimensional multivariate Gaussian (normal) distribution 
$\mathbf{X} = (X_1 \ \cdots \ X_s)^\top \sim \mathcal{N}(\boldsymbol{\mu}, \boldsymbol{\Sigma})$ 
for a mean vector $\boldsymbol{\mu} = (\mu_1 \ \cdots \ \mu_s)^\top$ and covariance matrix $\boldsymbol{\Sigma} = (\sigma_{ij})$. 
This can be implemented as
\begin{equation} \label{eqn:gibbs}
X_k \ \rvert \ \mathbf{X}_{-k} 
\sim \mathcal{N} (\boldsymbol{\mu}_k+ \boldsymbol{\Sigma}_{k, -k} \boldsymbol{\Sigma}_{-k, -k}^{-1}(\mathbf{X}_{-k}-\boldsymbol{\mu}_{-k}), 
\boldsymbol{\Sigma}_{k,k} - \boldsymbol{\Sigma}_{k, -k} \boldsymbol{\Sigma}_{-k, -k}^{-1} \boldsymbol{\Sigma}_{-k, k}),
\end{equation}
for $k = 1, \ldots, s$, which reduces to the iteration of the calculation of the one-dimensional normal distribution. \textcolor{red}{(Here, for simplicity of notation, the indices 
${}_k$ and ${}_{-k}$ represent the $k$th component and the components 
except for the $k$th component, respectively; e.g., 
$\boldsymbol{\Sigma}_{k, -k} = (\sigma_{k, 1}, \ldots, \sigma_{k, k-1}, \sigma_{k, k+1}, \ldots, \sigma_{k, s})$ and so on.)} Thus, we apply the inverse transform method in (\ref{eqn:gibbs}).
We set the parameter values $s = 3$ and 
\[ \boldsymbol{\mu} = \begin{pmatrix}
0\\
0\\
0
\end{pmatrix}, 
\boldsymbol{\Sigma} = \begin{pmatrix}
1 & 0.3 & -0.2 \\
0.3 & 1 & 0.5 \\
-0.2 & 0.5 & 1 
\end{pmatrix},\]
which were used in \cite[Chapter~6.1]{MR2710331}.

\begin{figure}
  \centering
\subfloat{
   \includegraphics[width=0.5\textwidth]{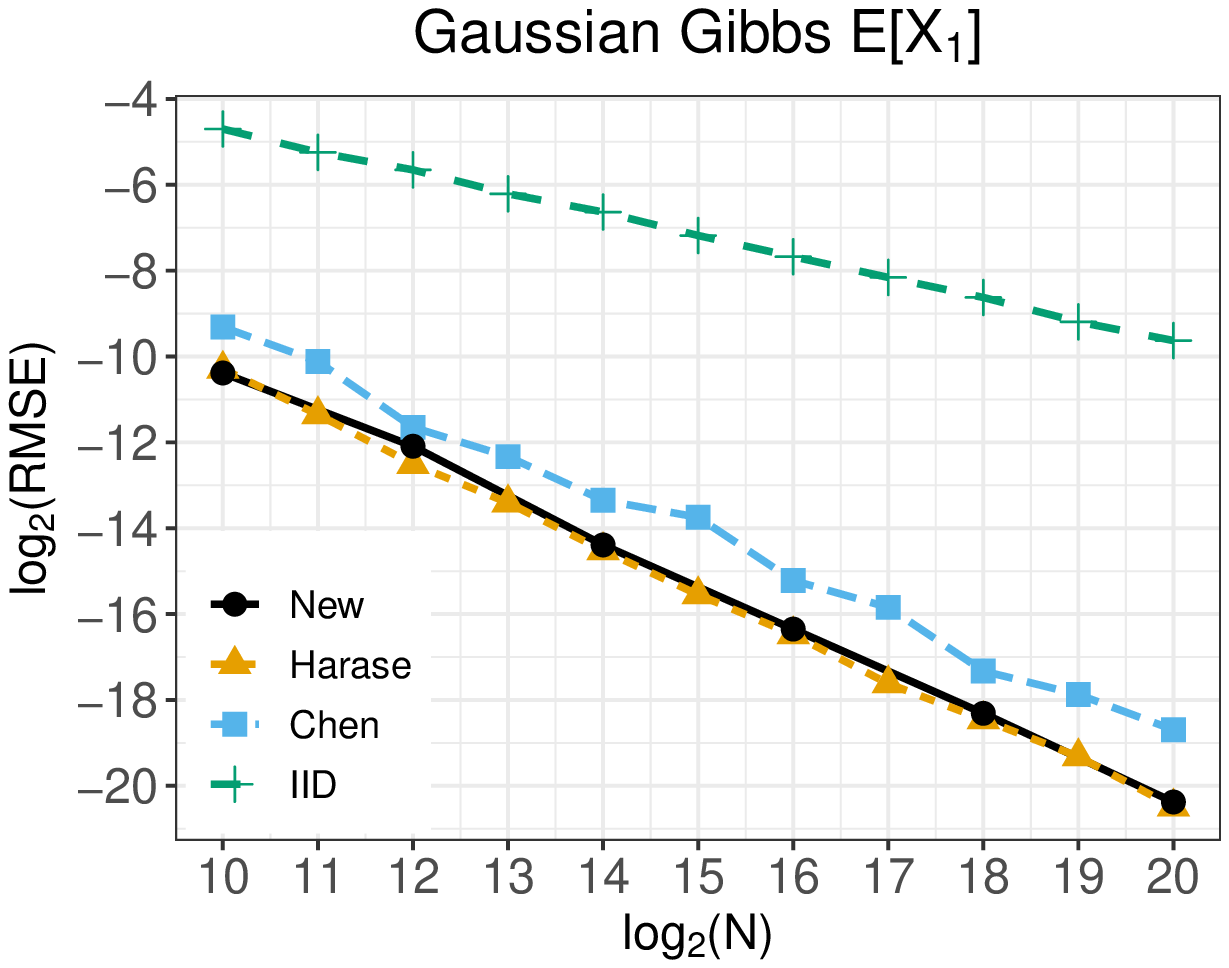}}\\
\subfloat{
   \includegraphics[width=0.5\textwidth]{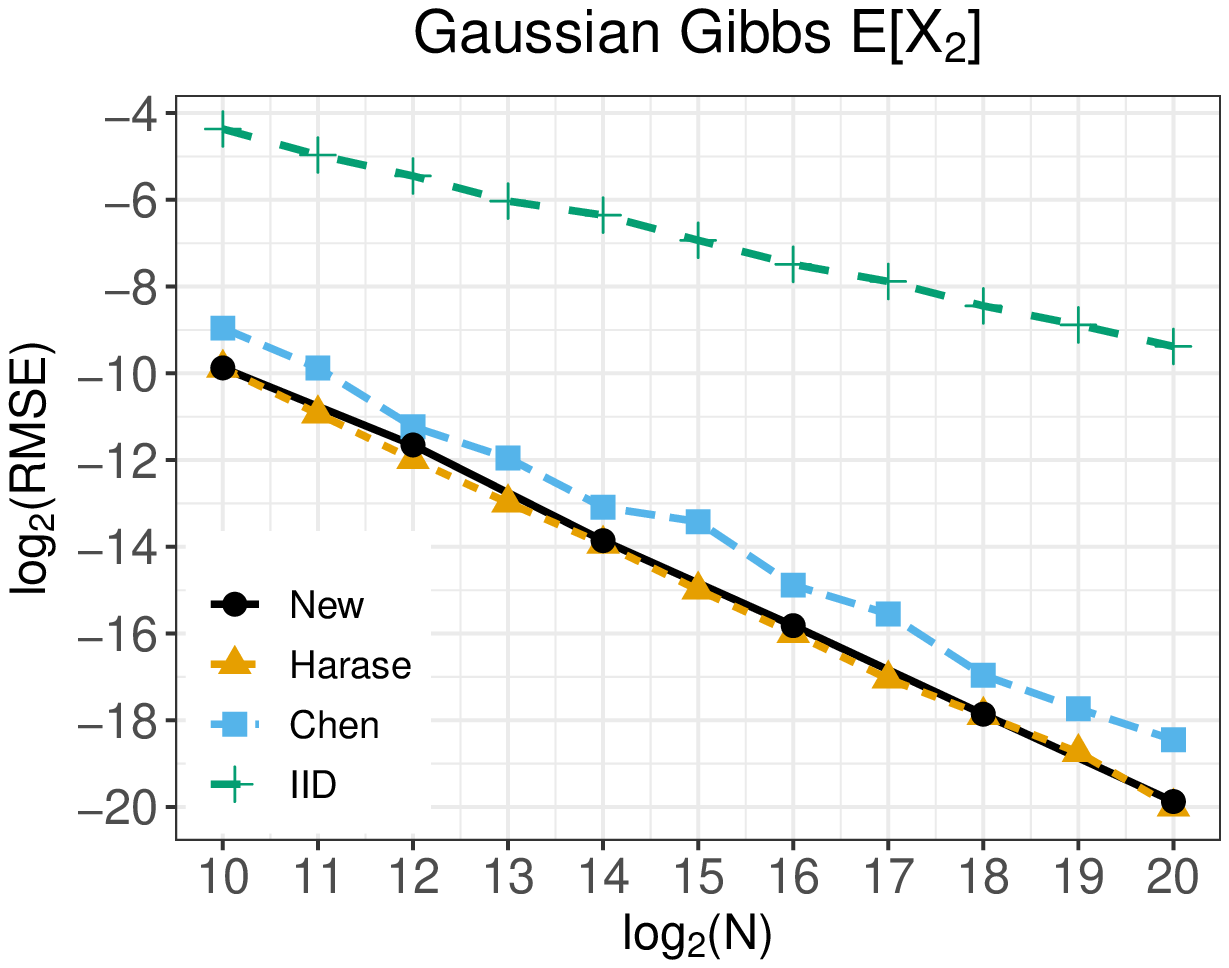}}~
\subfloat{
   \includegraphics[width=0.5\textwidth]{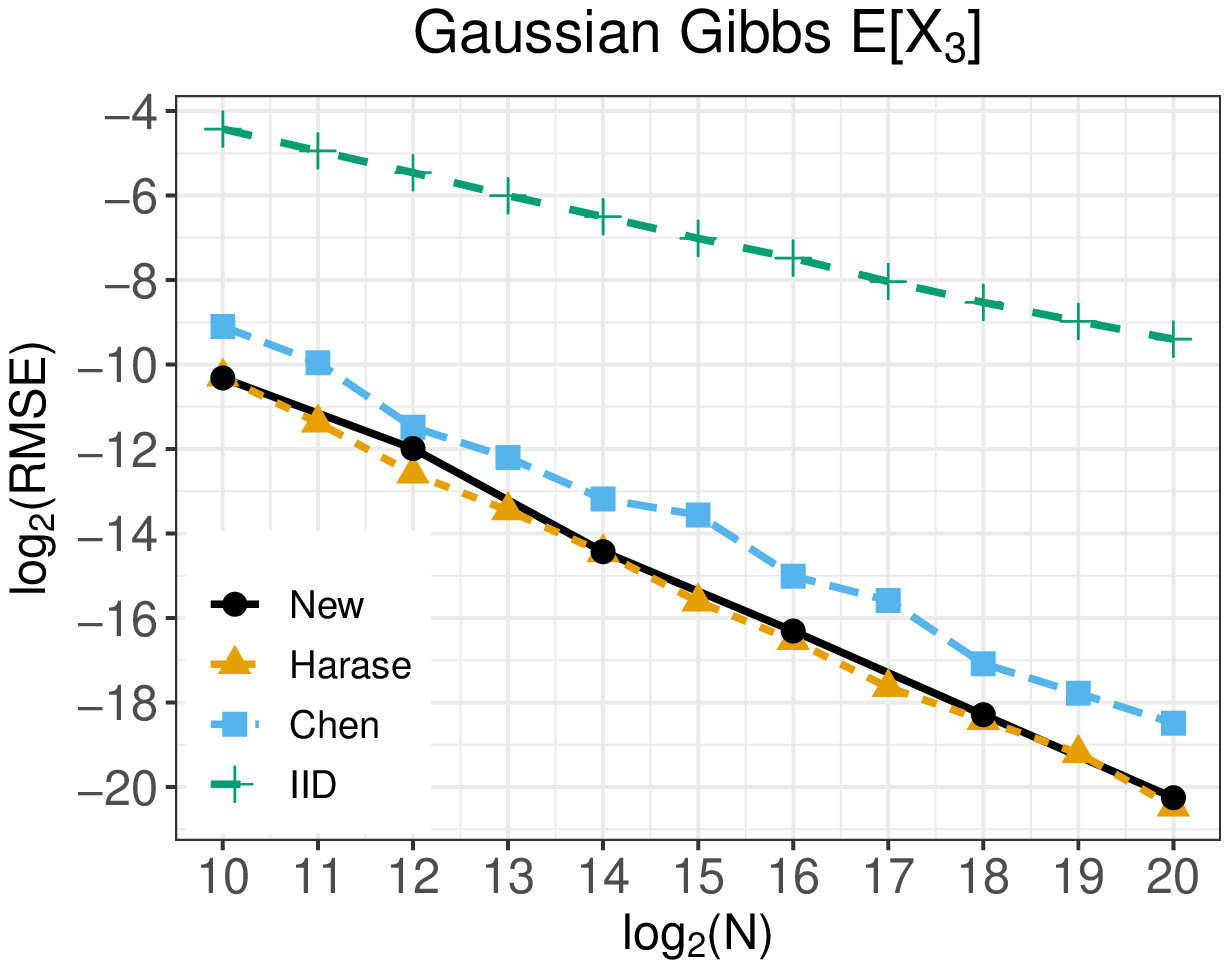}}
\caption{RMSEs for $E[X_1]$, $E[X_2]$, and $E[X_3]$ with true value $0$.}\label{fig:Gauss1}
\end{figure}

First, we estimate $E[X_1], E[X_2]$, and $E[X_3]$ with true value $0$ by taking the sample mean. 
Figure~\ref{fig:Gauss1} shows a summary of the root-mean-square errors (RMSEs) in $\log_2$ scale 
for sample sizes $N$ from $2^{10}$ to $2^{20}$ 
using 300 digital shifts. 
In all cases, the Tausworthe generators (labeled ``New'' and ``Harase'') 
optimized in terms of the $t$-value have almost the same accuracy and outperform Chen's generators. 

\begin{figure}
  \centering
  \subfloat{	
   \includegraphics[width=0.5\textwidth]{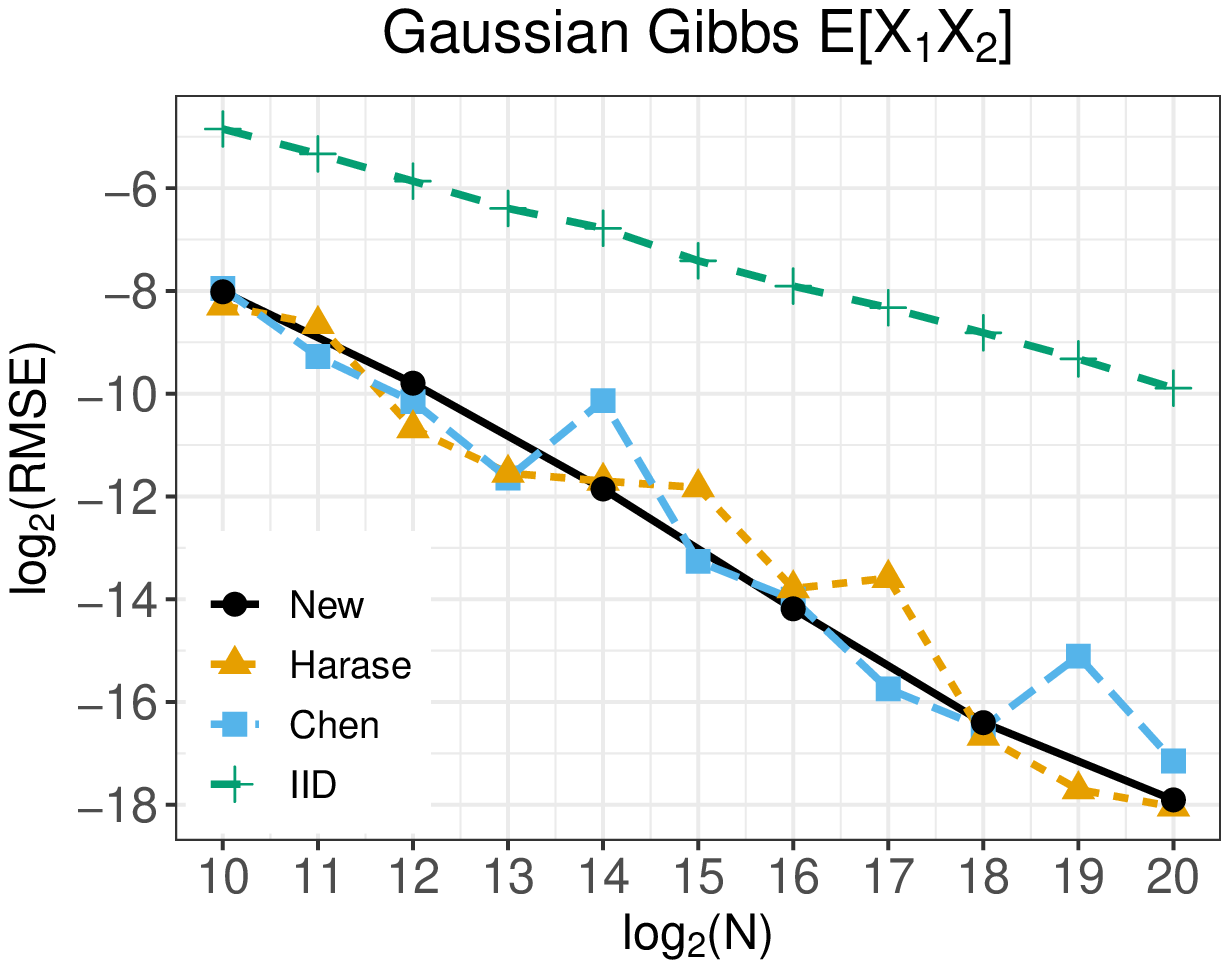}
  }\\ 
  \subfloat{	
   \includegraphics[width=0.5\textwidth]{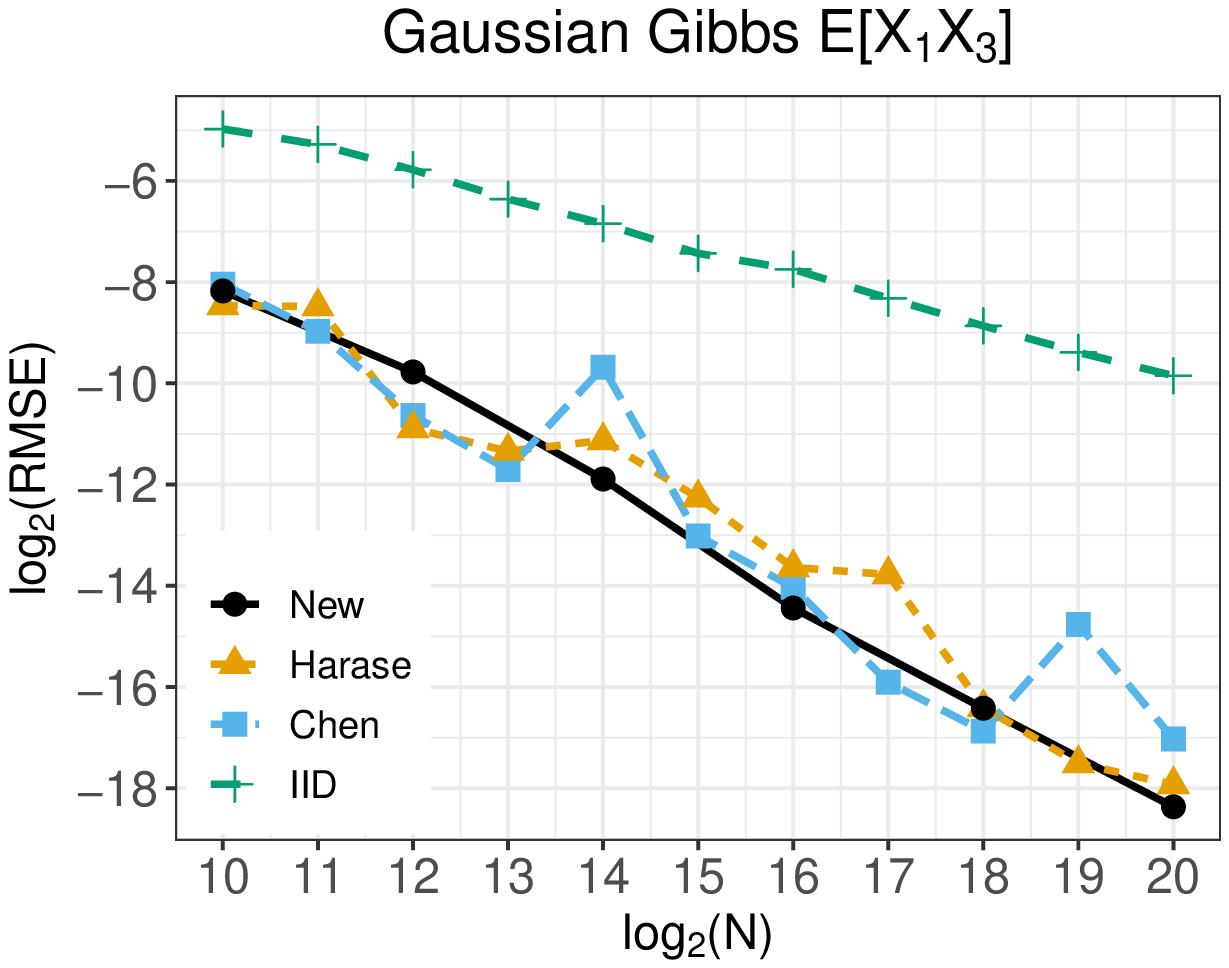}
  }~
  \subfloat{
   \includegraphics[width=0.5\textwidth]{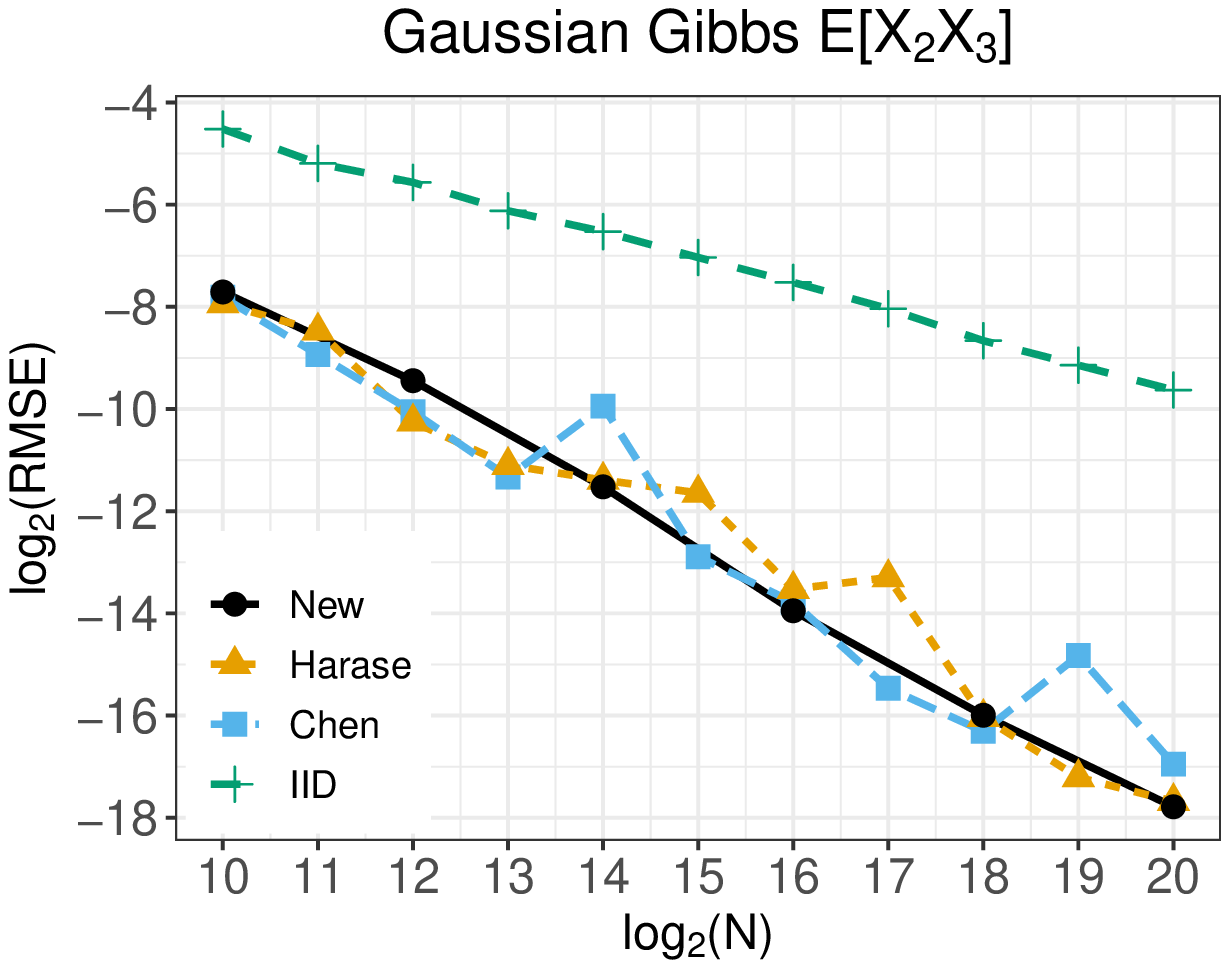}
  }
\caption{RMSEs for $E[X_1X_2]$, $E[X_1X_3]$, and $E[X_2X_3]$ with true values 0.3, -0.2, and  0.5.}
 \label{fig:Gauss2}
\end{figure}

\begin{figure}
\centering
\includegraphics[width=0.8\textwidth]{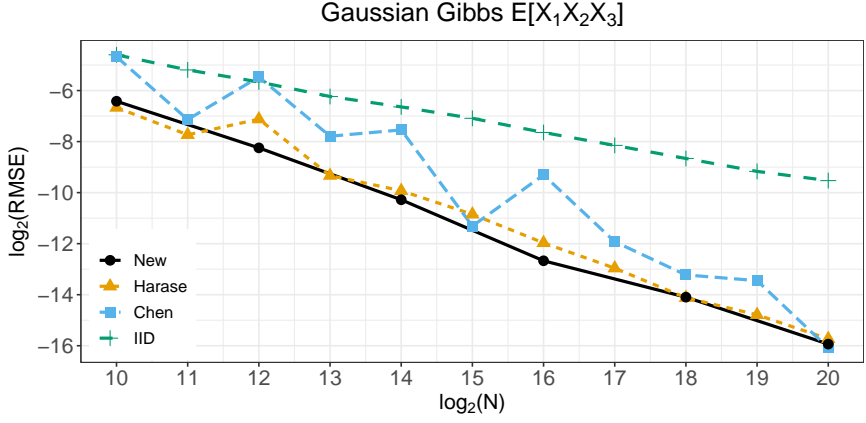}
\caption{RMSEs for $E[X_1 X_2 X_3]$ with true value $0$.}
\label{fig:Gauss3}
\end{figure}

Furthermore, we estimate the second-order moments $E[X_1X_2]$, $E[X_1X_3]$, 
$E[X_2X_3]$ and the third-order moment $E[X_1X_2X_3]$ using 300 digital shifts, respectively. 
Figures~\ref{fig:Gauss2} and \ref{fig:Gauss3} 
show summaries of the RMSEs. 
In Figure~\ref{fig:Gauss3}, we observe that Chen's generators are unstable and have several bumps when we estimate $E[X_1X_2X_3]$. 

\subsection{M/M/1 queuing system}

Our second example is an M/M/1 queuing model, which has the same setting 
as that in \cite[Chapter~8.3.2]{ChenThesis}. 
Consider a single-server queuing model, where the customers arrive 
as a Poisson process with intensity $\lambda > 0$ 
and the service time is exponentially distributed with intensity $\mu > 0$. 
Assume that $\mu > \lambda$ for system stability. 
Let $W_j$ denote the waiting time of the $j$th customer, $S_j$ denote the service time of the $j$th customer, 
and $T_j$ denote the time interval between the $j$th customer and the $(j-1)$th customer. 
Then, we have the Lindley recurrence: 
\begin{align}
& W_0 =  0, W_j  =  \max (W_{j-1} + S_{j-1} - T_{j}, 0), \label{eqn:queue1}\\
& S_{j-1}  \sim  \mathcal{E}\textit{xp}(\mu), T_{j} \sim \mathcal{E}\textit{xp}(\lambda), \label{eqn:queue2}
\end{align}
for $j \geq 1$, where $\mathcal{E}\textit{xp}(\cdot)$ denotes the exponential distribution. 
Under stationarity, 
the average waiting time is known as 
\begin{eqnarray} \label{eqn:queue3}
E[W_j] = \frac{\lambda}{\mu(\mu-\lambda)}
\end{eqnarray}
(cf. \cite{alma991001445919706711}).
We estimate the average waiting time (\ref{eqn:queue3}) by taking
the sample mean $(W_1 + \cdots + W_N)/N$ via Equations (\ref{eqn:queue1})--(\ref{eqn:queue2}). 
Note that we need $2N$ random points $(S_0 , T_1), \ldots, (S_{N-1}, T_N)$ for $N$ customers. 
Note also that the function $\max (\cdot, 0)$ is unsmooth at $0$. 

\begin{figure}[tbp]%
\centering
\includegraphics[width=0.8\textwidth]{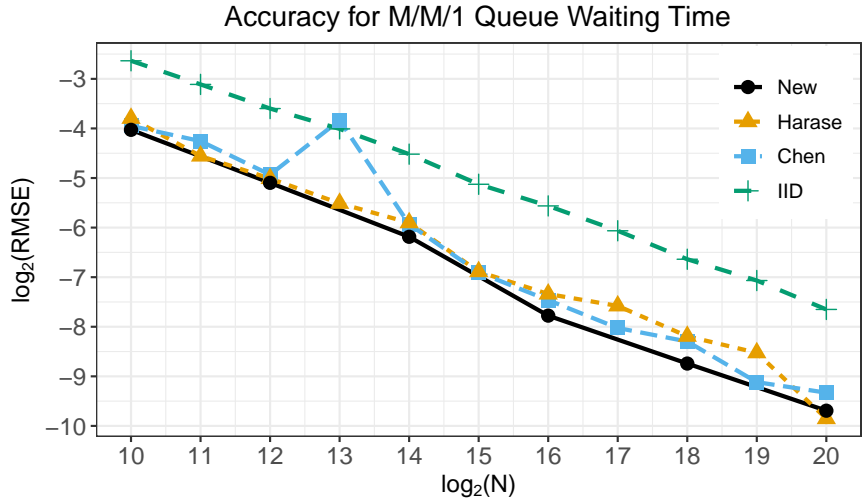}
\caption{RMSEs for the average waiting time of the M/M/1 queuing model.}\label{fig:queue}
\end{figure}

We set the parameters $\lambda=0.5$ and $\mu=1$. 
Figure~\ref{fig:queue} shows the RMSEs for the average waiting time using $300$ digital shifts. 
The three types of QMC points have almost the same performance, except for Chen's generator at 
$N = 2^{13}$. 
The bump of Chen's generator coincides with that in \cite[Figure~8.2]{ChenThesis}. 

\subsection{A linear regression model}

In the third example, we consider a linear regression model
\begin{eqnarray*} \label{eqn:lm}
y_i = \mathbf{x}_i^\top \boldsymbol\beta + \epsilon_i, \quad \epsilon_i \overset{\textrm{IID}}{\sim} \mathcal{N}(0, \tau^2) \quad (i = 1, \ldots, n),
\end{eqnarray*}
where $y_i$ is the $i$th observation on the response variable, 
$\mathbf{x}_i = (1, x_{i,1}, \ldots, x_{i, k})^\top$ is a $(k+1) \times 1$ vector of $1$ and $i$th observations on the $k$ explanatory variables, 
$\boldsymbol\beta = (\beta_0, \beta_1, \ldots, \beta_k)^\top$ is a $(k+1) \times 1$ vector of regression coefficients, and the error term $\epsilon_i$ is IID normal with mean zero and common variance $\tau^2$. 
Let $\textcolor{red}{\mathbf{X}} = (\mathbf{x}_1, \ldots, \mathbf{x}_n)^\top$ be an $n \times (k+1)$ design matrix (with rank $k+1 \textcolor{red}{< n}$) and $\mathbf{y} = (y_1, \ldots, y_n)^\top$ an $n \times 1$ vector. 

We now consider Bayesian inference as follows: 
We assume that the parameters $\boldsymbol\beta$ and $\tau^2$ are independent and have the 
\textcolor{red}{prior distributions} 
\begin{eqnarray*} 
\boldsymbol\beta & \sim & \mathcal{N}(\mathbf{b}_0, \textcolor{red}{\mathbf{B}}_0), \label{eqn:lm_prior1} \\
\tau^2 & \sim & \mathcal{IG} \left( \frac{n_0}{2}, \frac{s_0}{2}\right), \label{eqn:lm_prior2}
\end{eqnarray*}
where $\mathcal{IG}(\alpha_1, \alpha_2)$ denotes the inverse gamma distribution with shape parameter $\alpha_1 \textcolor{red}{>0}$ and \textcolor{red}{rate parameter} $\alpha_2 \textcolor{red}{>0}$, and 
\textcolor{red}{$(k+1)$-dimensional mean vector $\mathbf{b}_0$, $((k+1) \times (k+1))$-covariance matrix $\mathbf{B}_0$}, $n_0$, and $s_0$ are hyperparameters. 
Then, according to \cite{CHIB20013569,MR2648134}, sampling from the joint posterior distribution of 
$(\boldsymbol\beta^{\textcolor{red}{\top}}, \tau^2)^{\textcolor{red}{\top}}$ 
can be generated through sampling from the full conditional distributions 
\begin{eqnarray} \label{eqn:posterior}
\boldsymbol\beta \vert \tau^2, \mathbf{y} & \sim & \mathcal{N}(\mathbf{b}_1, \textcolor{red}{\mathbf{B}}_1), \label{eqn:lm_posterior1}\\
\tau^2 \vert \boldsymbol\beta, \mathbf{y} & \sim & \mathcal{IG} \left( \frac{n_1}{2}, \frac{s_1}{2}\right), 
\label{eqn:lm_posterior2}
\end{eqnarray}
where 
\begin{eqnarray*} 
\mathbf{b}_1 & = & \textcolor{red}{\mathbf{B}}_1 (\textcolor{red}{\mathbf{B}}_0^{-1} \mathbf{b}_0 + \tau^{-2}\textcolor{red}{\mathbf{X}}^\top \mathbf{y}), \textcolor{red}{\mathbf{B}}_1^{-1} 
= \textcolor{red}{\mathbf{B}}_0^{-1} + \tau^{-2} \textcolor{red}{\mathbf{X}}^\top \textcolor{red}{\mathbf{X}},\\
n_1 & = & n_0 + n, s_1 = s_0 + (\mathbf{y} - \textcolor{red}{\mathbf{X}} \boldsymbol\beta)^\top (\mathbf{y} - \textcolor{red}{\mathbf{X}} \boldsymbol\beta).
\end{eqnarray*}
Thus, we calculate $E[\boldsymbol\beta]$ and $E[\tau^2]$ by taking the sample mean 
using the Gibbs sampler based on (\ref{eqn:lm_posterior1}) and (\ref{eqn:lm_posterior2}).
We generate $\boldsymbol\beta$ in~(\ref{eqn:lm_posterior1}) via $\textcolor{red}{\mathbf{L}} (\Phi^{-1}(u_j), $ $\ldots, \Phi^{-1}(u_{j+k}))^\top$ for $u_j,  \ldots, u_{j+k} \in (0, 1)$, 
where $\textcolor{red}{\mathbf{B}}_1 = \textcolor{red}{\mathbf{L}}\textcolor{red}{\mathbf{L}}^\top$ is the Cholesky decomposition and $\Phi (\cdot)$ is
the cumulative distribution function of the standard normal distribution.

As a numerical example, we use the Boston housing data 
analyzed in \cite{HARRISON197881}. 
To investigate the demand for clean air, 
Harrison and Rubinfeld \cite{HARRISON197881} built a linear regression model given by
\begin{equation}
\begin{split} \label{eqn:boston}
\log (\textrm{MEDV}) & = \beta_0 + \beta_1 \textrm{CRIM} + \beta_2 \textrm{ZN} + \beta_3 \textrm{INDUS} + \beta_4 \textrm{CHAS} + \beta_5 \textrm{NOX}^2 \\
&   + \beta_6 \textrm{RM}^2 + \beta_7 \textrm{AGE} + \beta_8 \log(\textrm{DIS}) + \beta_9 \log (\textrm{RAD}) + \beta_{10} \textrm{TAX} \\ 
& + \beta_{11} \textrm{PTRATIO} + \beta_{12} \textrm{B} + \beta_{13} \log(\textrm{LSTAT}) + \epsilon, \quad 
\epsilon \sim \mathcal{N}(0, \tau^2), 
\end{split}
\end{equation}
where the housing price MEDV is a response variable, 
and $\textrm{CRIM}$, $\textrm{ZN}$, $\textrm{INDUS}$, $\textrm{CHAS}$, $\textrm{NOX}$, 
$\textrm{RM}$, $\textrm{AGE}$, $\textrm{DIS}$, $\textrm{RAD}$, $\textrm{TAX}$, 
$\textrm{PTRATIO}$, $\textrm{B}$, and $\textrm{LSTAT}$ are 13 explanatory variables 
(i.e., $k = 13$); 
see \cite[Table~IV]{HARRISON197881} for more details about the variables. 
In our experiment, we estimate the same linear regression model as in (\ref{eqn:boston}). 
Note that the state vector $(\boldsymbol\beta^{\textcolor{red}{\top}}, \tau^2)^{\textcolor{red}{\top}}$ has 15 dimensions (i.e., $s = 15$). 

\begin{figure}[tbp]%
\centering
\includegraphics[width=1\textwidth]{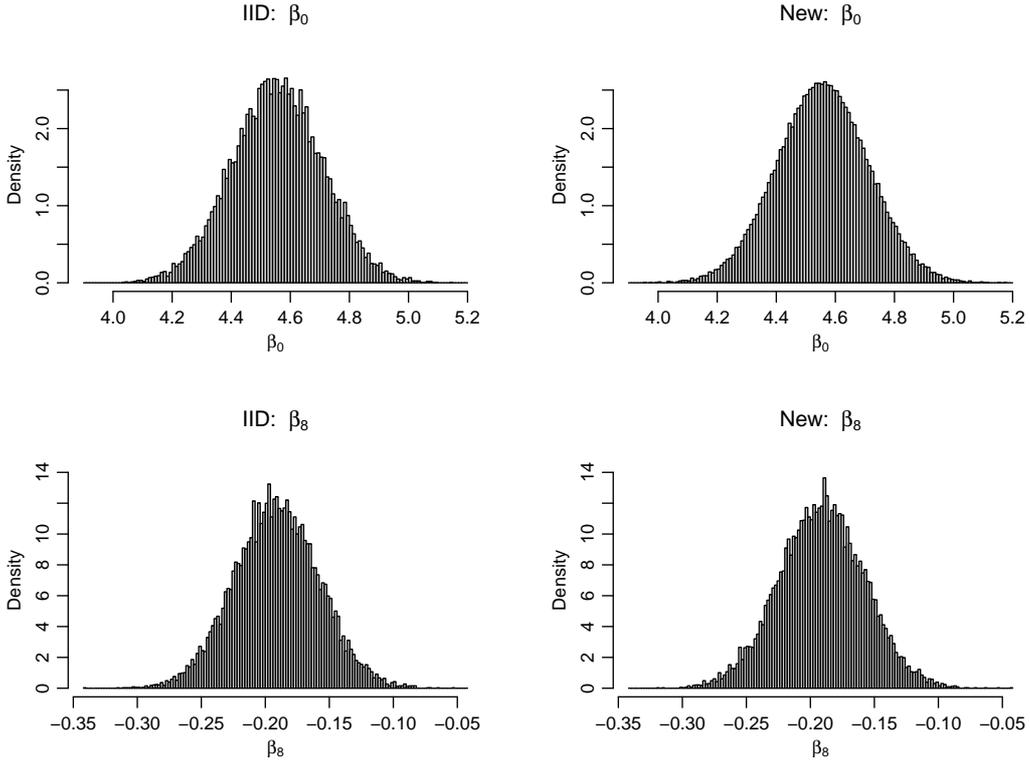}
\caption{\textcolor{red}{Histograms of $\boldsymbol\beta_0$ and $\boldsymbol\beta_8$ using $2^{14}$ IID uniform random points and QMC points generated by our new generator over $\mathbb{F}_4$.}}\label{fig:hist_beta}
\end{figure}

We set the hyperparameter values  
$\mathbf{b}_0 = \mathbf{0}$, $\textcolor{red}{\mathbf{B}}_0 = 100 \textcolor{red}{\mathbf{I}_s}$, $n_0 = 5$, $s_0 = 0.01$, 
and run the Gibbs sampler for $5000$ iterations using random numbers as a burn-in period. 
Then, we calculate $E[\boldsymbol\beta]$ and $E[\tau^2]$ by running 
the Gibbs sampler for $N = 2^{12}$, $2^{14}$, $2^{16}$, and $2^{18}$ iterations. 
Table~\ref{table:boston} shows a summary of sample variances of posterior mean estimates using 300 digital shifts. 
\textcolor{red}{
In both cases $\mathbb{F}_2$ and $\mathbb{F}_4$,  
Tausworthe generators optimized in terms of the $t$-value provide 
comparable to or better results than Chen's Tausworthe generators 
optimized in terms of the equidistribution property, excluding some exceptions 
(e.g., $\beta_6, \ldots, \beta_{13}$ for $N = 2^{14}$ 
estimated by Tausworthe generators over $\mathbb{F}_2$). 
Furthermore, we plot the histograms of $\beta_0$ and $\beta_8$ 
using $2^{14}$ IID uniform random points and QMC points generated by our new generator 
over $\mathbb{F}_4$ in Figure~\ref{fig:hist_beta}. 
In the case $\beta_0$, the sampling using our new QMC points tends to converge 
to the posterior distribution faster than the sampling using IID uniform random points, 
but in the case $\beta_8$, the difference seems to be unclear. 
From this, the estimation of $\beta_8$ 
might be more difficult than that of $\beta_0$ when we apply QMC.} 
Overall, our experiment implies that the $t$-value is a good measure of 
uniformity in the study of Markov chain QMC. 

\begin{table}
\caption{Variances of posterior mean estimates for $\boldsymbol\beta = (\beta_0, \ldots, \beta_{13})^{\textcolor{red}{\top}}$ and $\tau^2$.} \label{table:boston}
{\footnotesize
\textcolor{red}{
\begin{tabular}{|c|ccccccc|} \hline
\multicolumn{8}{|l|}{$N = 2^{12}$}\\ \hline
Parameter & $\beta_0$ & $\beta_1$ & $\beta_2$ & $\beta_3$ & $\beta_4$ & $\beta_5$ & $\beta_6$ \\ \hline
IID & 6.51e-06 & 3.56e-10 & 6.31e-11 & 1.26e-09 & 2.53e-07 & 3.28e-06 & 3.79e-10\\
Chen & 1.15e-09 & 6.29e-14 & 1.03e-14 & 2.52e-13 & 4.55e-11 & 5.38e-10 & 8.95e-14\\
Harase & 2.98e-10 & 1.41e-14 & 2.77e-15 & 5.60e-14 & 1.18e-11 & 1.27e-10 & 1.72e-14\\
New & 5.35e-10 & 3.59e-14 & 5.40e-15 & 1.40e-13 & 2.99e-11 & 3.85e-10 & 7.83e-14\\ \hline
\end{tabular} \\
\begin{tabular}{|c|cccccccc|} \hline
Parameter & $\beta_7$ & $\beta_8$ & $\beta_9$ & $\beta_{10}$ & $\beta_{11}$ & $\beta_{12}$ & $\beta_{13}$ & $\tau^2$\\ \hline
IID & 7.17e-11 & 2.92e-07 & 9.45e-08 & 3.60e-12 & 5.69e-09 & 3.07e-12 & 1.48e-07 & 1.22e-09\\
Chen & 1.37e-14 & 5.02e-11 & 1.60e-11 & 7.22e-16 & 1.18e-12 & 4.97e-16 & 3.19e-11 & 2.59e-13\\
Harase & 2.03e-15 & 1.18e-11 & 3.33e-12 & 1.59e-16 & 2.05e-13 & 1.16e-16 & 6.78e-12 & 1.27e-13\\
New & 7.26e-15 & 2.89e-11 & 1.06e-11 & 4.11e-16 & 6.26e-13 & 3.09e-16 & 2.09e-11 & 1.34e-13\\ \hline
\end{tabular}}
\textcolor{red}{
\begin{tabular}{|c|ccccccc|} \hline
\multicolumn{8}{|l|}{$N = 2^{14}$}\\ \hline
Parameter & $\beta_0$ & $\beta_1$ & $\beta_2$ & $\beta_3$ & $\beta_4$ & $\beta_5$ & $\beta_6$ \\ \hline
IID & 1.34e-06 & 1.17e-10 & 1.57e-11 & 3.13e-10 & 7.18e-08 & 7.61e-07 & 1.12e-10\\
Chen & 1.24e-10 & 9.66e-15 & 1.21e-15 & 3.27e-14 & 2.14e-11 & 6.62e-11 & 8.80e-15\\
Harase & 2.23e-11 & 1.75e-15 & 2.40e-16 & 5.19e-15 & 1.07e-12 & 1.26e-11 & 8.91e-15\\
New & 1.31e-11 & 6.83e-16 & 1.12e-16 & 4.55e-15 & 9.99e-13 & 1.09e-11 & 1.09e-15\\ \hline
\end{tabular} \\
\begin{tabular}{|c|cccccccc|} \hline
Parameter & $\beta_7$ & $\beta_8$ & $\beta_9$ & $\beta_{10}$ & $\beta_{11}$ & $\beta_{12}$ & $\beta_{13}$ & $\tau^2$\\ \hline
IID & 1.83e-11 & 6.36e-08 & 2.36e-08 & 9.15e-13 & 1.66e-09 & 6.75e-13 & 4.37e-08 & 3.00e-10\\
Chen & 1.46e-15 & 5.42e-12 & 1.81e-12 & 7.96e-17 & 1.29e-13 & 5.71e-17 & 3.32e-12 & 2.88e-14\\
Harase & 2.46e-15 & 6.10e-12 & 2.74e-12 & 9.44e-17 & 1.27e-13 & 7.52e-17 & 4.07e-12 & 2.34e-14\\
New & 1.29e-16 & 5.94e-13 & 1.69e-13 & 8.36e-18 & 1.26e-14 & 6.15e-18 & 3.43e-13 & 5.58e-15\\ \hline
\end{tabular}}

\textcolor{red}{
\begin{tabular}{|c|ccccccc|} \hline
\multicolumn{8}{|l|}{$N = 2^{16}$}\\ \hline
Parameter & $\beta_0$ & $\beta_1$ & $\beta_2$ & $\beta_3$ & $\beta_4$ & $\beta_5$ & $\beta_6$ \\ \hline
IID & 3.07e-07 & 2.74e-11 & 4.13e-12 & 8.43e-11 & 1.64e-08 & 1.72e-07 & 2.22e-11\\
Chen & 9.28e-12 & 6.59e-16 & 1.07e-16 & 2.37e-15 & 4.71e-13 & 4.73e-12 & 6.69e-16\\
Harase & 1.40e-12 & 9.12e-17 & 1.82e-17 & 3.55e-16 & 1.71e-13 & 7.11e-13 & 7.65e-17\\
New & 1.73e-12 & 9.86e-17 & 1.71e-17 & 4.20e-16 & 7.31e-14 & 9.01e-13 & 1.24e-16\\ \hline
\end{tabular} \\
\begin{tabular}{|c|cccccccc|} \hline
Parameter & $\beta_7$ & $\beta_8$ & $\beta_9$ & $\beta_{10}$ & $\beta_{11}$ & $\beta_{12}$ & $\beta_{13}$ & $\tau^2$\\ \hline
IID & 4.26e-12 & 1.54e-08 & 5.87e-09 & 2.25e-13 & 3.48e-10 & 1.89e-13 & 1.09e-08 & 7.35e-11\\
Chen & 1.14e-16 & 4.70e-13 & 1.67e-13 & 6.91e-18 & 1.08e-14 & 4.52e-18 & 2.34e-13 & 2.36e-15\\
Harase & 8.86e-18 & 4.47e-14 & 1.42e-14 & 1.18e-18 & 1.36e-15 & 6.42e-19 & 3.02e-14 & 9.24e-16\\
New & 1.63e-17 & 7.71e-14 & 3.20e-14 & 1.36e-18 & 2.28e-15 & 1.20e-18 & 4.68e-14 & 1.00e-15\\ \hline
\end{tabular}}

\textcolor{red}{
\begin{tabular}{|c|ccccccc|} \hline
\multicolumn{8}{|l|}{$N = 2^{18}$}\\ \hline
Parameter & $\beta_0$ & $\beta_1$ & $\beta_2$ & $\beta_3$ & $\beta_4$ & $\beta_5$ & $\beta_6$ \\ \hline
IID & 9.16e-08 & 5.65e-12 & 9.53e-13 & 1.85e-11 & 4.39e-09 & 5.28e-08 & 6.81e-12\\
Chen & 5.65e-13 & 3.33e-17 & 5.70e-18 & 1.36e-16 & 2.55e-14 & 2.73e-13 & 4.09e-17\\
Harase & 4.62e-14 & 2.48e-18 & 4.95e-19 & 1.07e-17 & 2.44e-15 & 2.40e-14 & 4.26e-18\\
New & 7.02e-14 & 5.55e-18 & 9.27e-19 & 2.10e-17 & 3.93e-15 & 4.46e-14 & 5.85e-18\\ \hline
\end{tabular} \\
\begin{tabular}{|c|cccccccc|} \hline
Parameter & $\beta_7$ & $\beta_8$ & $\beta_9$ & $\beta_{10}$ & $\beta_{11}$ & $\beta_{12}$ & $\beta_{13}$ & $\tau^2$\\ \hline
IID & 9.29e-13 & 3.75e-09 & 1.48e-09 & 5.92e-14 & 9.52e-11 & 3.47e-14 & 2.59e-09 & 1.52e-11\\
Chen & 6.90e-18 & 2.74e-14 & 9.00e-15 & 3.51e-19 & 5.77e-16 & 2.69e-19 & 1.56e-14 & 2.31e-16\\
Harase & 6.80e-19 & 2.49e-15 & 8.37e-16 & 3.47e-20 & 5.11e-17 & 2.51e-20 & 1.48e-15 & 3.08e-17\\
New & 8.56e-19 & 4.08e-15 & 1.19e-15 & 5.45e-20 & 7.38e-17 & 3.27e-20 & 1.89e-15 & 2.56e-17\\ \hline
\end{tabular}}
}
\end{table}

\begin{remark} \label{remark:ACF}
\textcolor{red}{
According to some heuristic arguments in \cite[Chapter~7.1]{MR2710331}, 
it is expected that Markov chain QMC drastically 
improves the rate of convergence 
when the dependence of states on the past decays quickly. 
To  investigate such phenomena, 
we plot the sample paths and autocorrelation functions (ACFs) of 
$\boldsymbol\beta_0, \boldsymbol\beta_1, \boldsymbol\beta_2$, and $\tau^2$ in Figure~\ref{fig:ACF} 
using $2^{14}$ IID uniform random points, after a burn-in period with 5000 iterations. 
The ACF plots imply that the dependence of states 
on the past decays very quickly (i.e., at one step) and has negligible effect on the current state.  
Moreover, it is believed 
that QMC methods in high-dimensional problems are successful 
especially in the case where the problems are dominated by the first leading variables 
or well approximated by a sum of functions of at most one or two variables (cf.~\cite{MR1966664}). 
Our linear regression example is probably included in such a class of problems, 
and hence, all the three Tausworthe generators drastically improve the rate of convergence. 
On the other hand, in more complicated applications in practice, the difference among these generators 
might not become clear, but we expect that Tausworthe generators optimized 
in terms of the $t$-value would be at worst superior to IID uniform random points. 
\begin{figure}[tbp]%
\centering
\includegraphics[width=1\textwidth]{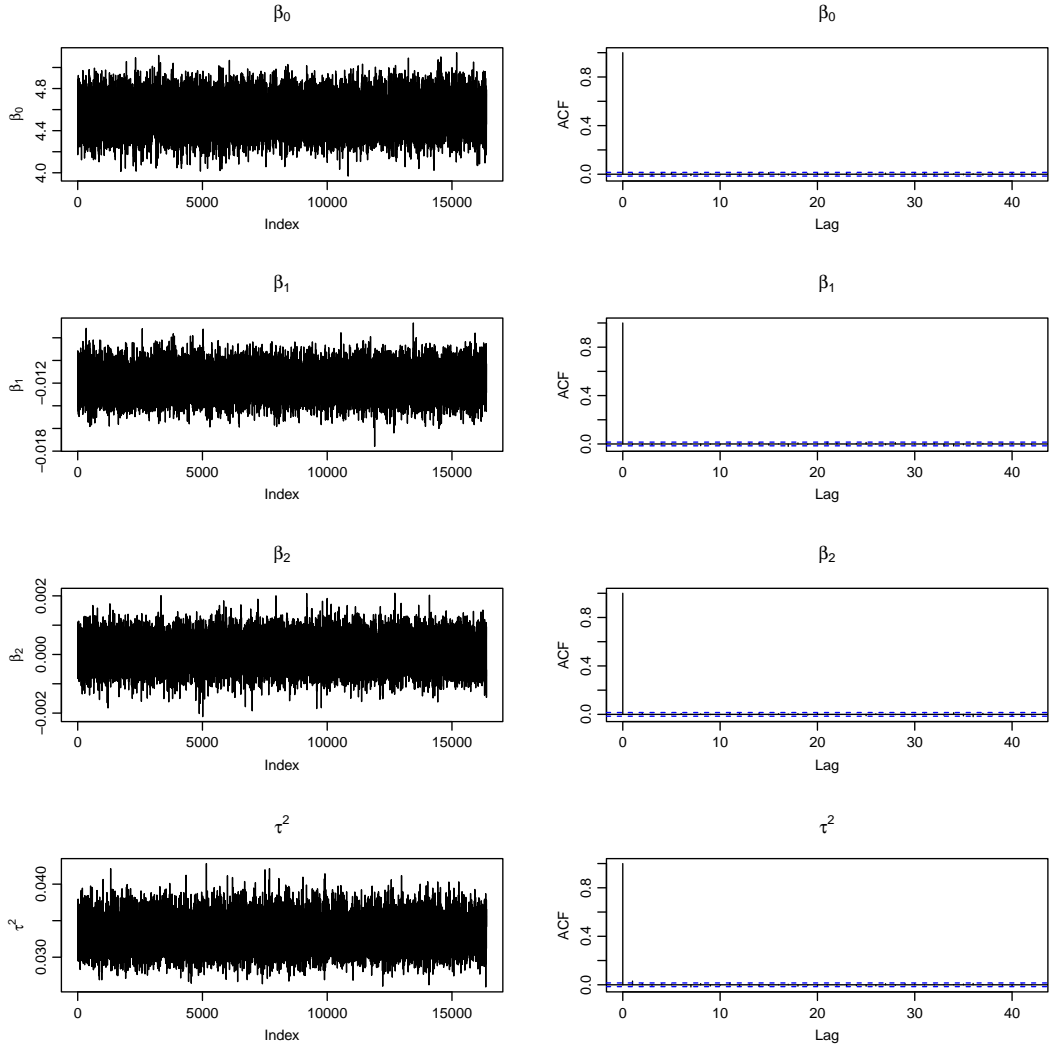}
\caption{\textcolor{red}{Sample paths and ACFs 
of $\boldsymbol\beta_0, \boldsymbol\beta_1, \boldsymbol\beta_2$, and $\tau^2$ 
using $2^{14}$ IID uniform random points.}}\label{fig:ACF}
\end{figure}
} 
\end{remark}

\begin{remark} \label{remark:skips}
The generation scheme in (\ref{eqn:blocks}) was originally used in \cite{MR2168266,MR2426105}. 
However, if $\textrm{gcd}(s, N-1) = d > 1$, 
then we have $d-1$ skips in (\ref{eqn:blocks}) through the entire period. 
To avoid such irregular skips, 
Tribble~\cite{MR2710331} and Chen~\cite{ChenThesis} suggested a strategy 
in which the skips are the same between every pair of non-overlapping $s$-blocks: 
Let $r$ be the smallest integer $r \geq s$ such that $\gcd (r, N-1) = 1$. 
Then, instead of (\ref{eqn:blocks}), 
we can consider $s$-dimensional non-overlapping points starting from the origin:
\begin{equation} \label{eqn:balance}
\begin{split}
(0, \ldots, 0),  & (u_{0}, \ldots, u_{s-1}) ,  (u_{r}, \ldots, u_{r+s-1}), \\
& (u_{2r}, \ldots, u_{2r+s-1}),  \ldots, (u_{\textcolor{red}{(N-2)r}}, \ldots, u_{\textcolor{red}{(N-2)r} + s-1}),
\end{split}
\end{equation}
which maintain balance (i.e., every $r$ steps) in each coordinate by discarding $r-s$ points between each block. 
We also implemented the strategy (\ref{eqn:balance}) and 
conducted the same experiments as in Section~\ref{sec:examples}. 
We obtained almost similar results with a slight fluctuation. 
In this paper, we optimized Tausworthe generators 
in terms of the $t$-value for consecutive output values, 
so we adopted the scheme (\ref{eqn:blocks}) without discarding $r-s$ points, 
which seems to be closer to the condition (\ref{eqn:Chentsov}). 
We refer the reader to \cite[Chapter~5]{MR2710331} and \cite[Chapter~8.2]{ChenThesis} for more details. 
\end{remark}

\section{Conclusion}\label{sec:conclusion}

We attempted to search for short-period Tausworthe generators over arbitrary finite fields $\mathbb{F}_b$ for Markov chain QMC in terms of the $t$-value. 
To achieve this, we generalized the search algorithms \cite{MR4143523,MR1160278} over $\mathbb{F}_2$ to those over $\mathbb{F}_b$. 
We conducted an exhaustive search, especially in the case where $b = 3, 4$, and $5$, 
and implemented Tausworthe generators over $\mathbb{F}_4$ with $t$-values zero for dimension $s = 3$, in addition to $s = 2$, and small for $s \geq 4$. 
We also reported numerical examples in which 
both Tausworthe generators over $\mathbb{F}_2$ and $\mathbb{F}_4$ 
optimized in terms of the $t$-value perform \textcolor{red}{comparable to or} better than Tausworthe generators 
\cite{MR3173841} optimized in terms of the equidistribution property. 

The two-element field $\mathbb{F}_2$ is the most important finite field in applications, 
but has some restrictions that do not occur over other fields $\mathbb{F}_b$. 
Therefore, in future work, it would be interesting to study implementations 
of other types of QMC points over $\mathbb{F}_b$, 
such as polynomial lattice rules \cite{MR2683394,MR1172997,Pirsic2001827} and 
irreducible Sobol'--Niderreiter sequences \cite{MR3926794}, which are closely related to the $t$-value. 
Furthermore, we are also plan\textcolor{red}{n}ing to apply our new generators\textcolor{red}{,} including \cite{MR4143523}\textcolor{red}{,} to a large variety of Bayesian computation using real-life data.

\textcolor{red}{
To conclude this paper, we mention some recent related works. 
In past a decade, the application of QMC methods to 
computational statistics has received a lot of attention 
for researchers and many novel studies have been proposed. 
For example, Chopin and Gerber \cite{MR3351446} 
present a class of algorithms where a sequential Monte Carlo strategy is implemented with QMC. 
Buchholz and Chopin \cite{MR3939383} derive approximate 
Bayesian computation (ABC) algorithms based on QMC 
for dealing with models with an intractable likelihood. 
As another direction, research on kernel density estimation using QMC 
has been actively conducted. We refer the reader to the survey paper \cite{10.1007/978-3-030-98319-2_1} for recent progress in this topic.}



\section*{Disclosure statement}
\textcolor{red}{No potential conflict of interest was reported by the author(s).}

\section*{Funding}
This work was supported by JSPS KAKENHI Grant Numbers JP22K11945, JP18K18016. 







\bibliographystyle{tfnlm}
\bibliography{harase_cud2}

\end{document}